\documentclass[a4paper,fleqn,usenatbib,useAMS]{mnras}

\usepackage{graphicx}	
\usepackage{amsmath}	
\usepackage{amssymb}	
\usepackage{multicol}        
\usepackage{bm}		
\usepackage{pdflscape}	

\usepackage[T1]{fontenc}
\usepackage{ae,aecompl}

\usepackage{newtxtext,newtxmath}

\title[From `bathtub' models to metallicity gradients]{From `bathtub' galaxy evolution models to metallicity gradients}
\author[F. Belfiore et al.]{F. Belfiore$^{1}$\thanks{ESO Fellow. E-mail: francesco.belfiore@eso.org} \& F. Vincenzo$^{2}$\thanks{E-mail: f.vincenzo@bham.ac.uk}, with R. Maiolino$^{3,4}$ and F. Matteucci$^{5,6,7}$.
\\
$^{1}$European Southern Observatory, Karl-Schwarzschild-Str. 2, Garching bei M{\"u}nchen, 85748, Germany\\
$^{2}$School of Physics and Astronomy, University of Birmingham, Edgbaston,  B15 2TT, UK\\
$^{3}$Cavendish Laboratory, University of Cambridge, 19 J. J. Thomson Avenue, Cambridge CB3 0HE, UK\\
$^{4}$Kavli Institute for Cosmology, University of Cambridge, Madingley Road, Cambridge CB3 0HA, UK\\
$^{5}$Dipartimento di Fisica, Sezione di Astronomia, Universit\`a di Trieste, via G.B. Tiepolo 11, 34100, Trieste, Italy\\
$^{6}$INAF, Osservatorio Astronomico di Trieste, via G.B. Tiepolo 11, 34100, Trieste, Italy\\
$^{7}$INFN, Sezione di Trieste, Via Valerio 2, 34100, Trieste, Italy}

\begin{document}

\date{Accepted Year Month Day. Received Year Month Day; in original form Year Month Day}

\pagerange{\pageref{firstpage}--\pageref{lastpage}} \pubyear{2016}

\maketitle

\label{firstpage}


\begin{abstract}
We model gas phase metallicity radial profiles of galaxies in the local Universe by building on the `bathtub' chemical evolution formalism - where a galaxy's gas content is determined by the interplay between inflow, star formation and outflows. In particular, we take into account inside-out disc growth and add physically-motivated prescriptions for radial gradients in star formation efficiency (SFE). 
We fit analytical models against the metallicity radial profiles of low-redshift star-forming galaxies in the mass range $\log(M_\star/M_\odot)$ = [9.0-11.0] derived by \cite{Belfiore2017a}, using data from the MaNGA survey. The models provide excellent fits to the data and are capable of reproducing the change in shape of the radial metallicity profiles, including the flattening observed in the centres of massive galaxies. We derive the posterior probability distribution functions for the model parameters and find significant degeneracies between them. The parameters describing the disc assembly timescale are not strongly constrained from the metallicity profiles, while useful constrains are obtained for the SFE (and its radial dependence) and the outflow loading factor. The inferred value for the SFE is in good agreement with observational determinations. The inferred outflow loading factor is found to decrease with stellar mass, going from nearly unity at $\log(M_\star/M_\odot) = 9.0$ to close to zero at $\log(M_\star/M_\odot) =11.0$, in general agreement with previous empirical determinations. These values are the lowest we can obtain for a physically-motivated choice of initial mass function and metallicity calibration. We explore alternative choices which produce larger loading factors at all masses, up to order unity at the high-mass end.
\end{abstract}




\section{Introduction} 
\label{sec:0}

Within the $\Lambda$ cold dark matter ($\Lambda$CDM) framework of galaxy formation, galaxy discs grow by cooling of baryonic gas at the centres of dark matter haloes \citep{Silk1977, White1978, White1991}. Gas is consumed by star formation and lost to the hot halo and the intergalactic medium via outflows driven by supernovae, stellar winds and radiation pressure \citep{Naab2017}. A detailed understanding of the processes driving this `baryon cycle' remains elusive, due to the difficulty of directly observing gas flows in and out of galaxies \citep{Sancisi2008, Almeida2014} and our limited understanding of the microphysics of the different feedback processes involved. Metals, which are direct products of stellar nucleosynthesis, represent ideal tracers of the baryon cycle. Studies of the metal content of galaxies may therefore be used to indirectly probe gas accretion and the effect of feedback mechanisms.

Observations of chemical abundances in external galaxies demonstrate the existence of a tight relation between luminosity, or stellar mass, and metallicity (the mass-metallicity relation, \citealt{Lequeux1979, Tremonti2004}). More recently evidence has accumulated in favour of the existence of a secondary dependence of the mass-metallicity relation on star formation rate (SFR, \citealt{Ellison2008, Mannucci2010, Lara-Lopez2010}). The observed correlation goes in the sense that galaxies of a fixed stellar mass have lower metallicity when they have higher SFR, or gas mass (\citealt{Hughes2013,Bothwell2013, Cresci2018a}, although see \citealt{Sanchez2018} for an alternative viewpoint).
 
These observations have motivated the development of chemical evolution models generally referred to as `gas regulatory' or `bathtub' models \citep{Bouche2010, Finlator2008, Dayal2013, Dekel2013, Lilly2013, Peng2014}. In this framework the traditional closed-box chemical evolution model \citep{Schmidt1963} is extended to take  inflows and outflows into account. The power of this approach lies in its simplicity and the ability to capture the basic physics behind galaxy scaling relations and/or abundance patterns of stars in the Milky Way \citep{Andrews2017, Weinberg2017}. 

Focusing on our Galaxy, there is a long history of chemical evolution models aimed at reproducing the metallicity gradient observed in the disc \citep{Lacey1985, Chiosi1980, Matteucci1986, Matteucci1989, Boissier1999, Chiappini2001}. In order to address the G-dwarf problem  \citep{Schmidt1963, Lynden-Bell1975_book}, these models generally assume continuous accretion of gas over Gyr timescales. Notably, the study of the metallicity gradient of the Milky Way has been instrumental in providing early support for the `inside-out' disc formation paradigm. In this framework, outer regions of the disc are formed later and on longer timescales, as expected from the theory of disc assembly in a cosmological context \citep{White1991, Kauffmann1996}.
Several classical models, however, reproduce the properties of the Milky Way without including outflows. The outcome is generally successful since the outflow loading factor is highly degenerate with the value of the nucleosynthetic yields, which are plagued by significant uncertainties \citep{Romano2010}. We note, moreover, that a radial dependence of the yield (which could be caused by radial changes in the initial mass function, IMF) or of the star formation efficiency (SFE= SFR/M$\rm _{gas}$, where M$\rm _{gas}$ is the cold gas mass) can also generate a negative metallicity gradient \citep{Goetz1992}. The effects of these different parameters (infall timescale, outflow loading factor, SFE etc) on the metallicity gradients have been discussed qualitatively in previous work, but the possible degeneracies between them remain difficult to quantify.

The advent of a new generation of large integral field spectroscopy (IFS) surveys (including CALIFA, \citealt{Sanchez2012}; and MaNGA, \citealt{Bundy2015}) has greatly improved the amount and quality of data relating to chemical abundances of external galaxies. Metallicity has been known to be a decreasing function of galactocentric distance in disc galaxies since the 1970s \citep{Searle1971, Peimbert1979, Shaver1983, Vila-Costas1992}, but modern IFS surveys have finally allowed studies of representative samples of galaxies with sufficient statistics to uncover more subtle trends. For example, \cite{Belfiore2017a} demonstrated that the shape of the gas-phase metallicity radial profiles depends on the stellar mass of the host galaxy. In particular, low-mass galaxies ($\rm log(M_\star/M_\odot) =9.0$) have flatter gradients than galaxies with stellar masses of $\rm log(M_\star/M_\odot)=10.5$. However, the metallicity radial profile is found to flatten again in the inner regions of the most massive star forming galaxies (see also \citealt{Sanchez-Menguiano2017a}). 

This mounting body of observational data justifies the development of chemical evolution models that may successfully cross the gap between integrated and spatially resolved properties of galaxies. Several attempts have already been made to use variants of the bathtub chemical evolution model to interpret resolved chemical abundances \citep{Ascasibar2015, Belfiore2015, Kudritzki2015, Ho2015, Lian2018}. 
In this paper we follow the same philosophy and develop an extension of the gas regulatory formalism in order to test whether simple analytical models can reproduce the observational trends highlighted in \cite{Belfiore2017a}. We focus entirely on gas-phase metallicity, and in particular on reproducing the detailed \textit{shape} of the radial metallicity profiles. We do not approximate radial profiles as linear gradients, since the observations are not well-represented by simple straight-line models. 

Our models take into account inside-out growth and radial variations of the SFE, but are otherwise intentionally simplistic. Metallicity gradients in real galaxies are likely to be affected by additional physics, which we do not include here (e.g. radial gas flows, the effect of enriched gas inflow or galaxy mergers). Zoom-in hydrodynamical simulations have been used to study these effects in some detail \citep{Torrey2012, Pilkington2012, Gibson2013, Tissera2018}. These simulations remain, however, too expensive to study large and representative samples of galaxies and explore variations in model parameters.

The flexible analytical models presented in this work, on the other hand, allow for a rapid exploration of parameter space, and are therefore ideally suited to developing physical intuition and uncovering the degeneracies between model parameters. We demonstrate the latter point explicitly by fitting our analytical models to the MaNGA radial metallicity profiles presented in \cite{Belfiore2017a}, and evaluating the model likelihood via Monte Carlo Markov chain (MCMC) sampling. 

In Sec. \ref{sec:1} we discuss the details of our chemical evolution model, including the prescription for inside-out growth and radial gradients in SFE. We also comment on the resulting time evolution of the metallicity gradient in the models and the effects of the four model parameters on the metallicity gradient. In Sec. \ref{sec:2} we describe our Bayesian fitting strategy and the observed degeneracies in the inferred best-fit models. In section \ref{sec:3} we discuss how the best-fit model parameters compare with  theoretical predictions, focusing specifically on the outflow loading factor.

%
\section{The chemical evolution model} 
\label{sec:1}

\begin{figure*} 
	\includegraphics[width=0.75\textwidth, trim=0 0 0 20, clip]{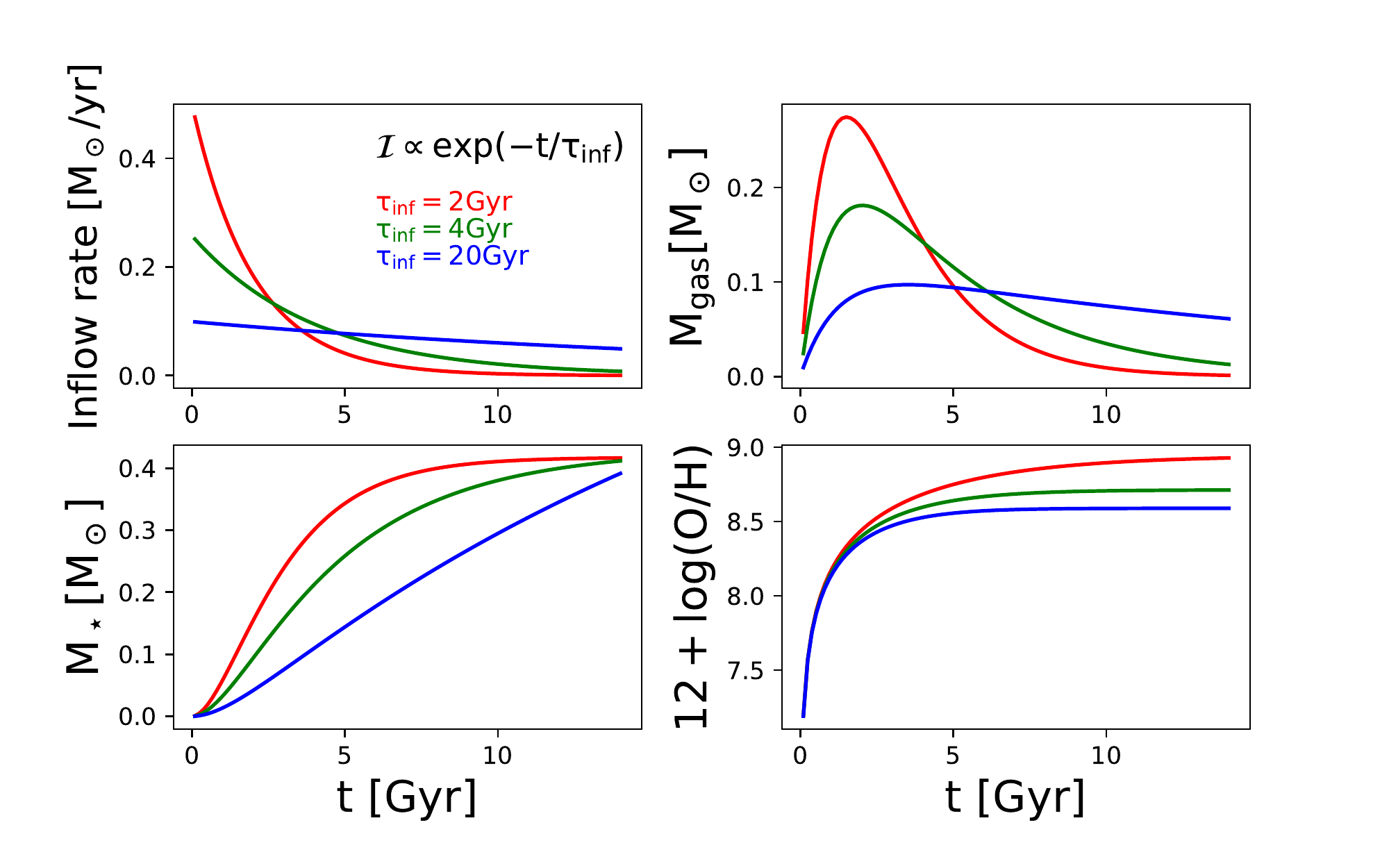}
	\caption{The time evolution of several physical parameters ($\rm \mathcal{I}, \rm M_{gas}, M_\star,  12+log(O/H)$) in our adopted chemical evolution model with exponentially decreasing inflow rate. Model tracks with different colours have different infall timescales. In this model, after an initial period of gas accumulation, where the inflow rate is larger than the SFR, the gas mass decreases with time and part of the mass in converted into stars. The total mass of the inflow is the same for all the models, leading to galaxies with very similar stellar masses at late times. The metallicity increases rapidly at early times but quickly reaches an equilibrium value.} 
	\label{fig1}
\end{figure*}

\subsection{The bathtub chemical evolution approach}
\label{sec:1.1}

In this work we make use of the bathtub chemical evolution model and consider a galaxy as a collection of independent radial annuli. Within each annulus we adopt the instantaneous recycling approximation \citep{Tinsley1980_book}. In this simplified framework one assumes that all stars more massive than $\rm M_{long-lived}$ (generally taken to be $\rm 1 \ M_\odot$) die instantaneously, and those of lower masses live forever. We further posit that the metals produced by the previous generation of stars are immediately and uniformly mixed with the pre-existing interstellar medium (ISM) of the region considered. Following standard notation we adopt the oxygen yield per stellar generation (y) and return fraction ($\mathcal{R}$) calculated by \cite{Vincenzo2016} using the \cite{Kroupa1993} IMF, respectively $\rm (y, \mathcal{R}) = (0.0105,0.285)$. For the rest of this work we characterise the yield normalising to the total mass of gas taking part in star formation, which we denote as p and is trivially related to y by $\rm y = p/(1-\mathcal{R})$.

Within the instantaneous recycling approximation, denoting the oxygen fraction (by mass) in the ISM as $\rm Z$, the gas mass as $\rm \Sigma_{g}$, the star formation rate as $\rm \Sigma_{SFR}$, the stellar mass as $\rm \Sigma_\star$, the outflow rate as $\mathcal{O}$ and the inflow rate as $\mathcal{I}$, each annulus is described by the following set of constitutive equations

\begin{equation}
\label{eq1}
\rm \frac{ \text{d} \Sigma_{ \star } }{ \mathrm{d} t } = (1- \mathcal{R}) \ \Sigma_{ \mathrm{SFR} } ;   
\end{equation}

\begin{equation}
\label{eq2}
\rm \frac{ \text{d} \Sigma_{ \mathrm{g} } }{ \mathrm{d} t } = - ( 1 - \mathcal{R}  ) \, \Sigma_{ \mathrm{SFR} } 
- \mathcal{O} + \mathcal{I};   
\end{equation}

\begin{equation}
\label{eq3}
\rm \frac{ \text{d} (Z \ \Sigma_{g}) }{ \mathrm{d} t } = p \  \Sigma_{ \mathrm{SFR} } - \mathcal{O} \ Z - 
( 1 - \mathcal{R}  ) \ Z \  \Sigma_{ \mathrm{SFR} } +\rm Z_{acc}  \mathcal{I},
\end{equation}
where $\rm Z_{acc}$ is the metallicity of the accreting gas. 

Together with Eqs. \ref{eq1}--\ref{eq3}, we will adopt the assumptions of the `ideal bathutb' model, namely:

\begin{enumerate}
	\item{A linear star formation relation, 
		\begin{equation}
		\rm \Sigma_{SFR} = \nu~ \Sigma_{g}, \label{SFlaw}
		\end{equation} 
		with star formation efficiency (SFE) $\nu$ which is constant in time.}
	\item{Outflow rate proportional to the SFR through a constant (in time) outflow loading factor ($\lambda$) 
		\begin{equation}
		\rm \mathcal{O} = \lambda \ \Sigma_{ \mathrm{SFR}}.
		\end{equation}}
\end{enumerate}

Combining equations \ref{eq1}-\ref{eq3}, we can re-write the time evolution of the gas-phase metallicity Z as

\begin{equation}
\rm \frac{dZ}{dt} =  \nu \left( p+ \frac{ (Z_{acc} - Z) \mathcal{I}}{SFR} \right)= p \ \nu - Z \frac{\mathcal{I}} {\Sigma_{g}},            \label{eqz}
\end{equation}
where to get to the final expression we have assumed that accretion is pristine ($\rm Z_{acc}  =0$). This assumption is adopted for the rest of this work.

\subsection{Time dependence of the inflow rate and resulting star formation history}
\label{sec:1.2}

Chemical evolution models aimed at reproducing the metallicity gradient in the Milky Way generally assume a faster assembly time for the inner disc, in order to mimic the theoretical expectation of inside-out growth \citep{Larson1976}. A simple way of implementing inside-out growth is to assume an exponentially declining accretion rate, with an infall timescale ($\rm \tau_{inf}$) which increases with galactocentric radius \citep{Chiosi1980, Matteucci1989, Boissier1999, Chiappini2001}
\begin{equation}
\label{eq6}
\rm \mathcal{I}(r,t) = \mathcal{I}_0(r) \ e^{-t/\tau_{inf}(r)}.
\end{equation}

This parametrization of the time dependence of the inflow rate is entirely predicated on its simplicity and does not have an a priori physical motivation. Other popular assumptions for the time evolution of the accretion rate include taking a constant inflow rate \citep{Peng2014}, using a redshift-dependent inflow rate which is proportional to the dark matter accretion rate computed in numerical simulations \citep{Forbes2014}, assuming the inflow rate necessary to reproduce the redshift dependence of the star formation main sequence \citep{Leitner2012, Lilly2016a} or assuming that the inflow rate is proportional to the SFR. The latter choice has been repeatedly used in recent literature \citep{Dayal2013, Kudritzki2015} because it simplifies the equations of chemical evolution, but prevents us from studying the non-equilibrium behavior of the system. In this work we consider the full time evolution of the solutions to the bath-tub model and not only their behaviour near equilibrium (i.e. $\rm d\Sigma_{g} / dt \sim 0$). We discuss some of the differences between these alternative assumptions and their relation to equilibrium in Appendix \ref{sec:appA}.


Assuming  eq. \ref{eq6} and integrating eq. \ref{eq1} - \ref{eq3} with respect to time, we derive analytical time-dependent solutions for this chemical evolution model given in Table \ref{table_sol}. Since the constitutive equations of our model are first order ordinary differential equations, the solutions are trivially obtained by standard methods and we omit the derivation.\footnote{Some details regarding the derivation of our analytical solutions are presented in the Appendix \ref{sec:appA}. Recently \cite{Spitoni2017} and \cite{Weinberg2017} have presented similar analytical solutions and further details on their derivation. Notably, the \cite{Spitoni2017} solution are identical to the ones derived in this work, after taking the difference in notation into account. The solutions in \cite{Weinberg2017} are slightly different because the authors regard star formation history, and not the mass accretion rate, as fundamental, and the star formation history obtained in our model does not match exactly any of the simple cases discussed in their paper.} 
The solutions have four free parameters: the star formation efficiency $\rm \nu$, the outflow loading factor $\rm \lambda$ , the inflow timescale $\rm \tau_{inf}$ and the normalisation of the inflow rate $\rm \mathcal{I}_0$. 

\begin{table}
	\caption{Exact analytical time-dependent solution of the exponential infall models used in this paper. The relevant timescales as the equilibrium timescale $\rm \tau_{eq} \equiv \frac{1}{\nu  (1-\mathcal{R} + \lambda)}$ and the critical timescale $\rm \tau_{c}^{-1} \equiv \tau_{eq}^{-1}- \tau_{inf}^{-1} $.}	
	\footnotesize
	
	\begin{tabular}{ l c}
		\hline
		Galaxy property 	& 	Solution \\
		\hline
		
		$\rm \Sigma_{g} $ & $ \rm  \mathcal{I}_0~e^{-t/\tau_{inf}}~\tau_c~(1-e^{-t/\tau_c}) $	\\
		$\rm \Sigma_{SFR} = \nu~\Sigma_{g} $ & 	
		$\rm  \nu~\mathcal{I}_0~e^{-t/\tau_{inf}}~\tau_c~(1-e^{-t/\tau_c})$ \\
		$\rm \Sigma_{\star}$  & 		$ \rm (1-\mathcal{R}) \nu \tau_c \mathcal{I}_0 \left(  \tau_{inf} (1-e^{-t/\tau_{inf}}) - \tau_{eq} (1-e^{-t/\tau_{eq}}) \right)$ \\
		$\rm Z_{g}$ & 	$ \rm p \nu \left(  \tau_c - t \frac{e^{-t/\tau_c}}{1-e^{-t/\tau_c}}  \right) $ \\
			
	\end{tabular}	
	\label{table_sol}	
\end{table} 

In Fig. \ref{fig1} we show the time evolution of different physical quantities (the inflow rate, gas mass, stellar mass and metallicity) in this model, considering three different inflow timescales ($\rm \tau_{inf} = 2, 4, 20 ~Gyr$). The inflow rate is normalised so that the total mass of gas accreted between $t=0$ and 14 Gyr is the same in each model, and is set to one in arbitrary units.
The other parameters are held fixes at $\rm (\varepsilon, \lambda)= (0.5 ~Gyr^{-1}, 1.0)$. Because of this normalisation, the total stellar mass at late times is nearly the same between different models, with small differences due to the different final gas masses and fraction of mass expelled due to outflows. 

Since SFR $\rm \propto ~ M_{gas}$ in our model, the time evolution of the gas fraction tracks the star formation history (SFH) of the system. Exponential infall models produce SFHs similar to the popular $\rm SFR(t) \propto t ~e^{-t/\tau}$ `delayed SFH' parameterization, which is in reasonable agreement with the mean SFH obtained by inverting the star formation main sequence \citep{Leitner2012, Ciesla2017} and expectations from simulations \citep{Simha2014}. In our models the gas mass increases linearly with time at early times, as gas accumulates in the system faster than it can be processed. At late times, on the other hand, the SFH follows the exponential decay of the inflow rate, as star formation is limited by the available gas supply.

The history of chemical enrichment in these models can also be divided into two phases. During the first phase metallicity increases rapidly as the metals quickly pollute the pristine gas and SFRs are high. At late times, however, the metallicity reaches an equilibrium value, since chemical enrichment is balanced by metal consumption from star formation and expulsion by outflows. This stage of `chemical equilibrium' at late times is a general feature of gas regulatory models which include inflows and outflows \citep{Peng2014, Weinberg2017}. In our models the equilibrium abundance depends on the three parameters $\rm (\nu, \lambda, \tau_{inf})$. As can be seen from Fig. \ref{fig1}, longer infall timescales correspond to more extended SFH and lower equilibrium metallicities. A more detailed discussion of the equilibrium solutions and their significance in our models is presented in Appendix \ref{sec:appA}.

\subsection{Radial dependence of model parameters}
\label{sec:1.3}

\subsubsection{Infall rate and timescale}
\label{sec:1.3:1}
In this work we use the parametrisation of the infall timescale adopted for the Milky Way models of \cite{Matteucci1989} and subsequent revisions thereof. This formalism adopts an infall timescale which increases linearly with radius,

\begin{equation}
\rm	\tau_{inf}(r) = a + b \ r,
\end{equation}
where $a$ is the infall timescale at the galaxy centre (r=0) and $b$ represents the linear gradient of the infall timescale. A positive $b$ is required to mimic inside-out growth. The choice of this particular functional form is again dictated by its simplicity and is only weakly constrained post-facto by its ability to fit current abundance and gas fraction data in our Galaxy. We note that even when radial flows are explicitly modelled, the inclusion of radial flows does not lead to inside-out growth, but simply to a different `effective accretion' profile (see the discussion in e.g. \citealt{Pezzulli2016}). 

In addition to the infall timescale, our model depends on the normalisation of the accretion rate profile $\rm \mathcal{I}_0(r)$. This normalising factor must have a radial dependence in order to generate a negative gradient in the stellar mass surface density.  While most models assume $\rm \mathcal{I}_0(r) = A ~ \exp(-r/h)$, where $h$ is a scale-length determined by fitting to the data, no simple prescription for the radial dependence of $\rm \mathcal{I}_0(r)$ is capable of generating a disc which remains exponential at all times. However, for suitably high values of $a$ and $b$ this standard choice of the normalisation parameter generates discs which are roughly exponential, especially at large radii. 

Fortunately, the normalisation of the inflow rate only has an effect on \textit{extensive} quantities (like $\rm M_{gas}$ or $\rm M_\star$), but not on quantities which are ratios of the above (like sSFR=$\rm SFR/M_\star$, $\rm f_{gas}$ or 12+log(O/H)). We can demonstrate this explicitly in the case of metallicity by noting the the solution in Table \ref{table_sol} does not depend on $\rm \mathcal{I}_0$. 
\footnote{Note that this is only true for a linear star formation law of the form of eq. \ref{SFlaw} and is not the case if one assumes the relation between SFR and gas mass to be given by a power law of the type $\rm SFR \propto M_{gas}^k$.} In this work we only fit the metallicity gradient and therefore do not consider the normalisation of the inflow rate as a free parameter.


\subsubsection{The star formation law and efficiency}
\label{sec:1.3:2}
\begin{figure} 
	
	\includegraphics[width=0.49\textwidth, trim=0 20 0 20, clip]{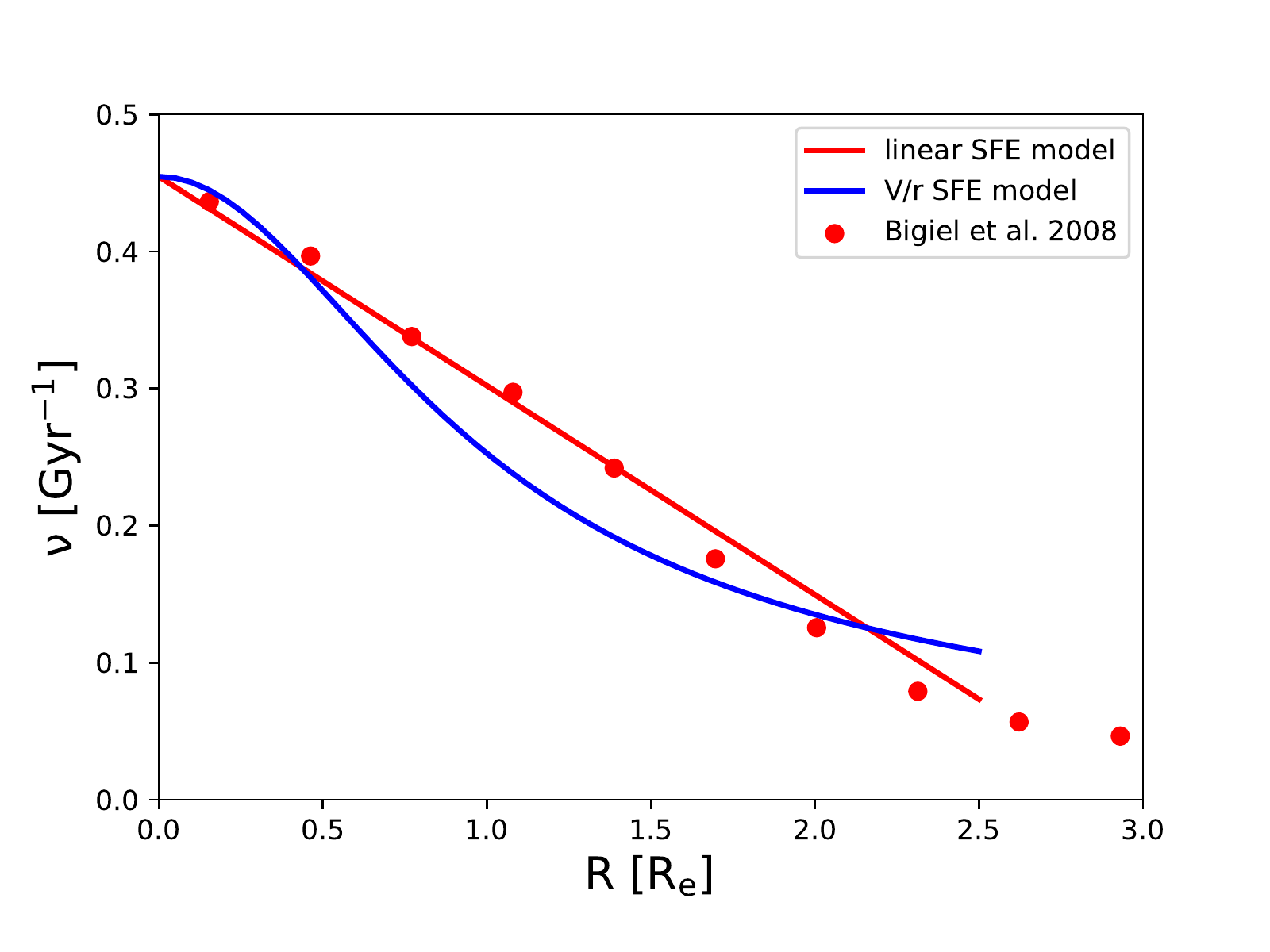}
	\caption{The star formation efficiency ($\rm \nu$) as a function of radius for different models considered in this work. The blue curve corresponds to the assumption $\rm \nu \propto V/r$. The rotation curve is parametrised using a tanh model, leading to the SFE expression presented in eq. \ref{SFE1}. The red curve shows an alternative model where the SFE decreases linearly with $\rm r/R_e$. This model is a good representation of the observed SFE radial gradient in nearby galaxies (red data points, from \protect\citealt{Bigiel2008}). For illustration purposes the red curve shown in figure represents the best fit to the observational data, and the blue curve is fixed to have the same SFE at $\rm r=0$.}
	\label{fig2}
\end{figure}


\begin{table*}
	\caption{Free parameters in the adopted chemical evolution model. (1) and (2) correspond to the two different models for the radial variation of the SFE, discussed in Sec. \ref{sec:1.3:2}.}
	\centering
	\begin{tabular}{ c c l c}
		\hline 
		parameter & definition & defining relation & unit\\
		\hline
		$a$  & infall timescale at galaxy centre & $\rm \tau_{inf} = a + b \ r$ & Gyr  \\
		$b$ & gradient of the infall timescale & $\rm \tau_{inf} = a + b \ r$  & $\rm Gyr \ kpc^{-1}$  \\
		(1) $\nu_0$  & free parameter in SFE ($\nu$) & (1) $\rm \nu=\nu_0 ~h/r~tanh(r/h) $ & $\rm Gyr^{-1}$   \\
		(2) $\nu_0'$ & "   & (2) $\rm \nu= 0.45  - \nu_0' r/R_e$ &  "  \\
		$\lambda$ & outflow loading factor & $\rm \lambda=\mathcal{O}/SFR$ & dimensionless \\
		\hline	
	\end{tabular}
	\label{table_param}
\end{table*}

\begin{figure*}
	\includegraphics[width=0.6\textwidth, trim=0 0 0 0, clip]{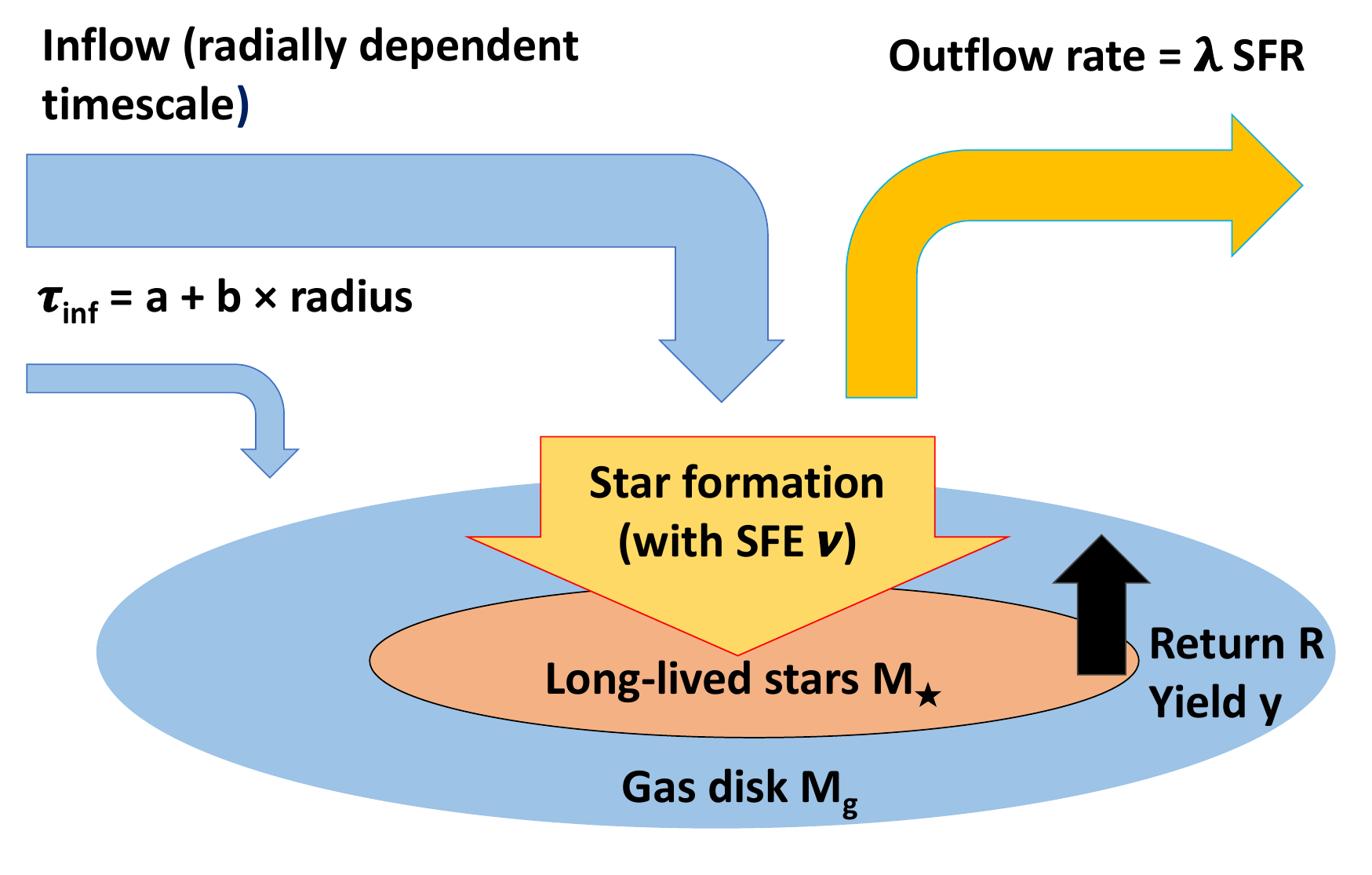}
	\caption{An illustration of the main components of the chemical evolution model described in this work. At the core of the model lie the equations of the `bathtub' or `gas regulatory' model, where star formation is regulated by inflow of pristine gas, the star formation efficiency of the disc gas ($\nu$) and star formation driven outflows (with loading factor $\lambda$). In order to mimic the inside-out growth of the disc, the infall timescale is assumed to be function of radius, with timescale $\rm \tau_{inf} = a + b~r$. Inner regions of the galaxy therefore form both earlier and faster. As described in Table \protect\ref{table_param}, we consider two different models for the radial variation of the star formation efficiency.}
	\label{fig2b}
\end{figure*}

While star formation is most directly associated to the molecular phase of the ISM \citep{Kennicutt2012}, for the purposes of chemical evolution the relevant gas mass to consider is the \textit{total} mass of gas diluting the metals. We assume that this consists of both atomic and molecular gas, with negligible mass in the ionised phase. Since atomic and molecular gas have different radial profiles, with the molecular gas being more centrally concentrated \citep{Leroy2009, Bigiel2012}, our model must take into account the different radial profiles of the star forming and the total gas component. In this work we parametrise this by using a linear star formation law (Eq. \ref{SFlaw}), and a star formation efficiency which decreases with radius.\footnote{A super-linear star formation law could also be used to a similar end, but this would prevent us from generating analytical solutions to the constitutive equations of the bathtub model, and is therefore not considered here.}

We consider two alternative parametrisations of the star formation efficiency and its radial dependence. The first is based on the orbital timescale, while the second assumes a fixed SFE for molecular gas and a radially decreasing molecular gas fraction. In the following we describe these models in more detail.

\begin{enumerate}
	\item{
	A classical implementation of the formation law is obtained assuming that the depletion time is proportional to the orbital timescale \citep{Silk1997, Kennicutt1998}. Under this assumption one may write the star formation efficiency as
	\begin{equation}
	\label{SFlaw1}
	\rm \nu \propto \tau_{orbit}^{-1} \propto \frac{V(r)}{r}, 
	\end{equation}
	where V(r) is the rotational velocity at radius r. 
	
	The rotation curve is a direct observable, which can be derived from the MaNGA data, but for simplicity we use a hyperbolic tangent model to represent a galaxy's rotation curve. While not strictly physically motivated, this model is found to reproduce the shapes of rotation curves of local galaxies to high accuracy \citep{Andersen2013, Westfall2014}. We fix the scale length of the rising part of the rotation curve to be the same as the exponential disc scale length, which is close to the relation observed to hold in local galaxies \cite{Amorisco2010}. Our model for the SFE is therefore given by  
	\begin{equation}
	 \label{SFE1}
	\rm \nu = \nu_0 ~h/r~tanh(r/h),
	\end{equation}
	where h is the exponential disc scale length and $\nu_0$ is a free parameter of the model, and corresponds to the star formation efficiency at r$=$0. In Fig. \ref{fig2} we show the SFE as a function of radius for a model galaxy using this SFE parametrization (solid blue line). In the rest of the paper we refer to this model as the $\rm \nu \propto V/r$ model.
	
} 
	\item{
	If the SFE depends on the free fall time of individual molecular clouds, and the mass spectrum of giant molecular clouds is roughly independent of the galactic environment the clouds live in, then we expect SFR $\propto$ $\rm M_{H2}$. This model is broadly supported by observations of the gas content of local galaxies on kpc-scales \citep{Bigiel2008, Leroy2009}. The ability of the ISM to form a molecular component is likely driven by the hydrostatic gas pressure and the interstellar radiation field. Analytical recipes exist to compute these quantities based on other observables \citep{Elmegreen1993, Blitz2006, McKee2010}. In this work, however, we wish to use the simplest possible prescription motivated by available data. To this end we fit the radial dependence of the SFE from \cite{Bigiel2008} with a linear model, which is found to be a good representation of the data. In particular, we assume $\rm SFR/M_{H2} = 2.0~ Gyr $ and take the radial dependence of $\rm M_{H2}/M_{HI}$ from Fig. 13 of \cite{Bigiel2008}. The redial dependence of the SFE is therefore determined entirely by the radial variation in $\rm M_{H2}/M_{HI}$.
	
	In Fig. \ref{fig2} we show the SFE from the \cite{Bigiel2008} data (red dots) and our best linear fit. In order to convert the $\rm R_{25}$ (radius where the galaxies reaches 25th magnitude in r-band) values used in \cite{Bigiel2008} to an effective radius ($\rm R_e$), we assume the galaxies in the \cite{Bigiel2008} sample to be exponential discs with canonical central surface brightness of 21.65 magnitude arcsec$^{-2}$ \citep{Freeman1970}.
	In this model the SFE is therefore set to vary linearly with radius according to 
	\begin{equation}
	\label{SFE2}
	\rm \nu = \nu_{0,c}  - \nu_0' r/R_e. 
	\end{equation}
	The best-fit parameters obtained fitting the \cite{Bigiel2008} data are  $\rm \nu_{0,c}=0.45 \ Gyr^{-1}$ and  $\rm \nu_0'= 0.15 \ Gyr^{-1}$. 
	
	While this SFE model has two free parameters, $\rm \nu_{0,c}$ and $\rm \nu_0'$, in this work we take $\rm \nu_0'$ as a free parameter and fix $\rm \nu_{0,c}$ to its best-fit value from the Bigiel et al. data. This choice is motivated by the desire to keep only one free parameter in the star formation law and by the fact that fixing $\rm \nu_0'$ generates a radial SFE dependence very similar to the $\rm \nu \propto V/r$ star formation model (eq. \ref{SFE1}, see Fig. \ref{fig2}). 
	By using a model with free $\rm \nu_0'$, on the other hand, we are able to test whether real metallicity gradients are best described by a steeper or shallower (or even flat) SFE gradient.
	The second SFE parametrization used in this paper is therefore given by equation \ref{SFE2} with $\rm \nu_{0,c}=0.45 \ Gyr^{-1}$.
	In the following we refer to this as the linearly decreasing SFE model.
}

\end{enumerate}

\subsubsection{The outflow loading factor}
\label{sec:1.3:3}

The outflow loading factor may be expected to depend on galactocentric distance, increasing towards the galaxy outskirts, where the local escape velocity may be lower. In this work, however, we refrain from introducing an ill-characterised radial dependence for the outflow lading factor and assume it to be constant as a function of radius. 

\subsubsection{Summary of the model}
\label{sec:1.3:4}

In summary, in this work we fit metallicity gradients with two sets of chemical evolution models, which differ in their treatment of the radial dependence of the SFE (the $\rm \nu \propto V/r$ model and the linearly decreasing SFE model). Both models have four free parameters, whose definitions and units are summarised in Table \ref{table_param}. In Fig. \ref{fig2b} we present a graphical summary of the main features of our chemical evolution framework, which highlights the similarities to the bathtub model of \cite{Lilly2013}. 

\begin{figure*} 
	\includegraphics[width=0.75\textwidth, trim=20 0 40 0, clip]{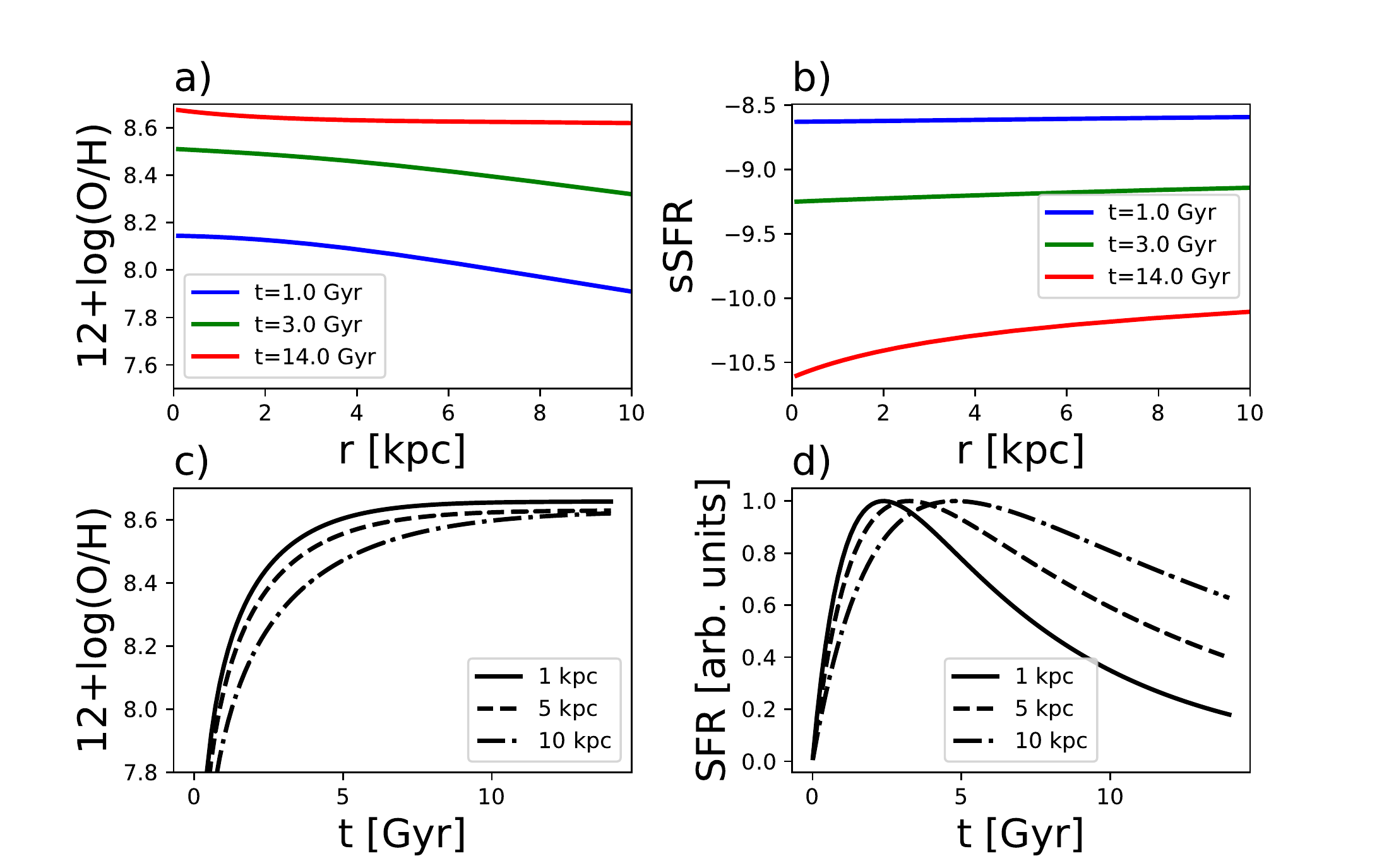}
	\caption{\textbf{a)} The time evolution of the metallicity radial profile using the $\rm \nu \propto V/r$ model and parameter values $\rm (a,b,\nu_0, \lambda)=(5.0,2.0,0.5, 1.0)$. \textbf{b)} Time evolution of the sSFR radial profile for the same model. \textbf{c)} Time evolution of the ISM metallicity at three different galactocentric distances, as noted in the legend.  \textbf{d)} The star formation histories of three regions at different galactocentric distances. Each SFH is normalised to its maximum value. The panel demonstrates out prescription for inside-out growth, where outer regions form stars later and more slowly.
	}
	\label{fig3}
\end{figure*}

\subsection{Time evolution of the metallicity gradient}
\label{sec:1.4}

In this subsection we explore the time evolution of the metallicity gradient and other physical quantities predicted by our chemical evolution model and the effect of varying its free parameters. We focus on the model with $\rm \nu \propto V/r$ but similar trends are obtained by studying the linearly decreasing SFE model. 

In Fig. \ref{fig3}a we show the metallicity gradient at three different times (t=1.0, 3.0, 14.0 Gyr) for the $\rm \nu \propto V/r$ model and example values of model parameters ($a$; $b$; $\nu_0$; $\lambda$)  = (5 Gyr; 1 Gyr/kpc; 0.5 $\rm Gyr^{-1}$; 1.0). While these parameter values have been chosen to be approximately representative of real galaxies, they are only used here for illustrative purposes.

Fig. \ref{fig3}a demonstrates that our model generally predicts a \textit{flattening} of the metallicity gradient over time. Panel c of Fig. \ref{fig3} shows that metallicity increases quickly at early times at all radii, with larger radii taking longer to reach the equilibrium metallicity value, as already noted in Sec. \ref{sec:1.2}. 

In Fig. \ref{fig3}b we show the sSFR radial profiles predicted by our chemical evolution model. At early times sSFR is high, and the radial profile is nearly flat. As the system evolves, however, the sSFR profile develops a dip at small galactocentric radii, indicative of gas exhaustion in the central regions of the galaxy. While in this work we do not fit observed sSFR profiles, we note that the shapes of the sSFR gradients produced by our chemical evolution model are at least qualitatively consistent with those observed in local \citep{Spindler2018, Belfiore2018} and high redshift \citep{Wang2017, Tacchella2018} galaxies.

Finally, in Fig. \ref{fig3}d we show the SFHs of galactic regions at different galactocentric distances. The SFHs have been normalised to their maximum value. The figure demonstrates the inside-out growth prescription embedded in our model, where regions at larger radii have both delayed (i.e. the peaks SFR occurs at later times) and more extended SFHs.

\begin{figure*} 
	
	\includegraphics[width=0.85\textwidth, trim=0 0 0 0, clip]{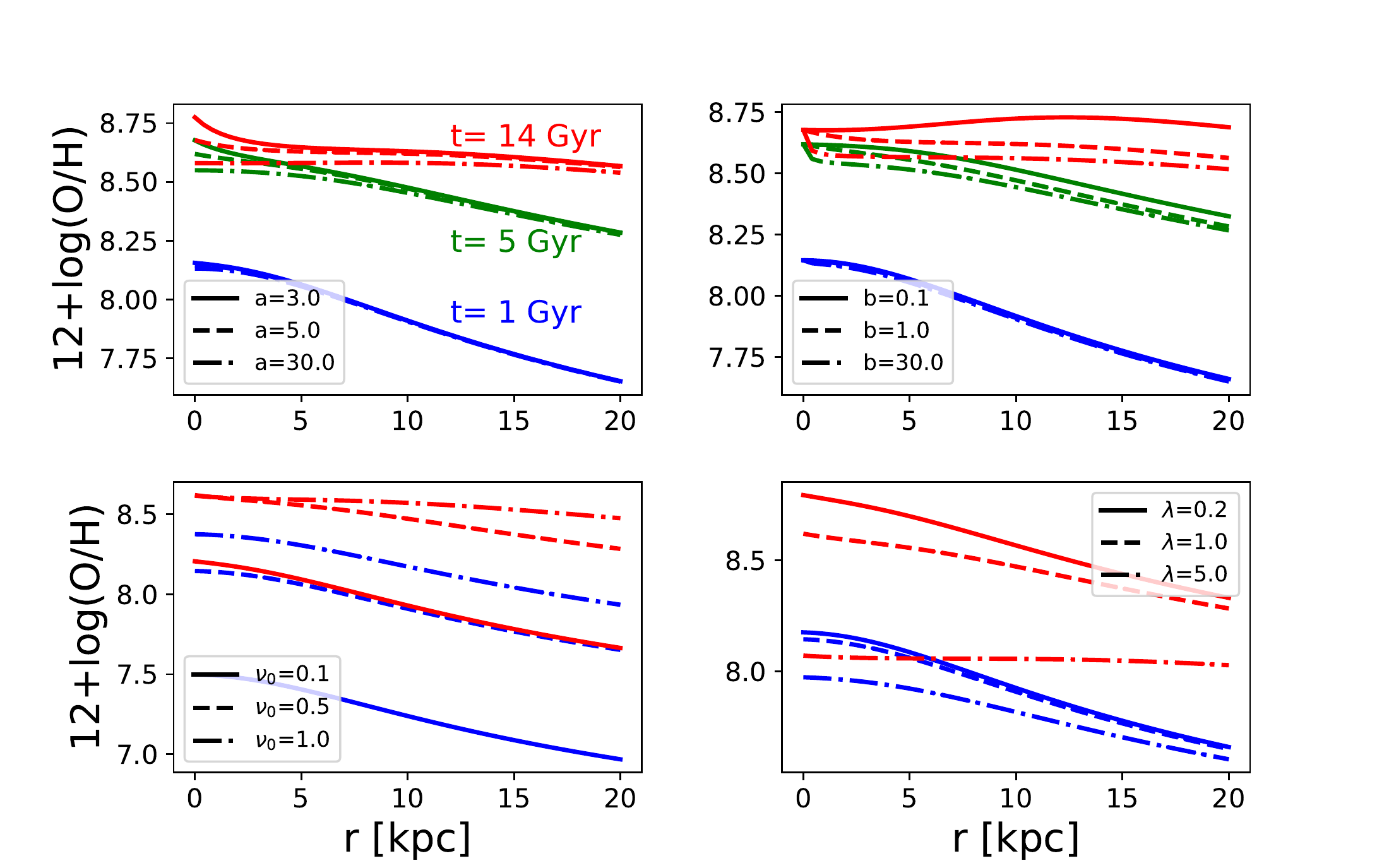}	
	\caption{The time evolution of the radial metallicity gradient for the $\rm \nu \propto V/r$ model. Blue, green and red curves in all panels refers to model predictions at t=1, 3, 14 Gyr respectively. Green curves are omitted in the bottom panels for clarity. In each panel a different model parameter is modified according to the legend. The fiducial model parameters are $\rm (a;b;\nu_0; \lambda)=(5.0 \ Gyr; 1.0 \ Gyr/kpc; 0.5 \ Gyr^{-1}; 1.0)$.}
	\label{fig4}
\end{figure*}

We next discuss the effect of the four free parameters  $a$, $b$, $\nu_0$, $\lambda$ on the time evolution of the metallicity gradient. In Fig. \ref{fig4}  we show the metallicity gradient at different times (1.0, 4.0 and 14.0 Gyr). In each panel one of the model parameters is varied and the other ones are kept fixed. The models shown in dashed lines correspond to the parameters values ($a$; $b$; $\nu_0$; $\lambda$)  = (5 Gyr; 1 Gyr/kpc; 0.5 $\rm Gyr^{-1}$; 1.) and are the same in all four panels. The main findings from this analysis are summarised below.

\begin{enumerate}
	\item{The $a$ parameter only affects the metallicity of the central region at late times. A small value of $a$ corresponds to a shorter infall timescale for the central regions and, therefore, a higher metallicity in the centre.}
	\item{The $b$ parameter mostly affects the slope of the metallicity gradient at late times. At early times the metallicity is mostly driven by the ability of the system to process the gas (via star formation and outflows), so the $b$ parameter does not affect the early evolution of the metallicity gradient. At late times, however, metallicity is driven by the availability of gas, which in turn depends on the inflow rate. The $b$ parameter, therefore, determines how fast the metallicity of outer regions catches up with the inner regions. A large value of $b$ implies a larger difference in the infall timescale between the centre and the outskirts and, therefore, a steeper metallicity gradient.}
	\item{The value of the SFE at r $=$ 0 ($\nu_0$) determines how fast gas can be processed into stars and, therefore, mostly impacts the early evolution of the system. Higher $\nu_0$ implies faster star formation and consequent enrichment, thus leading to higher metallicity at early times. At late times, the inner regions have already reached their equilibrium metallicity. A high $\nu_0$ therefore mostly affects the outer regions, allowing them to experience sufficient star formation to be chemically enriched. High $\nu_0$ at late times therefore corresponds to flatter gradients, which are the natural state of evolved systems in this model.}
	\item{The outflow loading factor $\lambda$ strongly affects the shape of the metallicity gradient and the maximum metallicity reached by the galaxy at both early and late times, although its impact is most significant at late times. A higher outflow loading factor leads to lower metallicities, as a larger fraction of the gas reservoir is expelled, therefore preventing further chemical enrichment. A larger loading factor also produces flatter gradients, as the outflow expels gas from the high-SFR central regions, preventing their early-time enrichment.}
\end{enumerate}


\section{Fitting the low-redshift metallicity gradients} 
\label{sec:2}

\subsection{Metallicity gradients in the nearby Universe}
\label{sec:2:1}

In this work we fit the metallicity radial profiles of star forming galaxies derived by \cite{Belfiore2017a}, making use of 550 galaxies from the MaNGA survey \citep{Bundy2015, Yan2016a}, part of SDSS-IV \citep{Blanton2017}. \cite{Belfiore2017a} find a mild change in the slope of the metallicity gradient as a function of mass, with low-mass galaxies having flatter gradients. Their data also shows a flattening and/or metallicity drop in the centres of massive galaxies ($\rm log(M_\star/M_\odot)>10.5$). More recently, \cite{Sanchez-Menguiano2017a} confirmed the presence of an inner drop in the metallicity gradient for massive galaxies using a sample of 102 galaxies observed at higher spatial resolution with the MUSE integral field spectrograph on the ESO Very Large Telescope. Albeit offering lower spatial resolution, the MaNGA dataset is unique in being representative of the local population of star forming galaxies in the stellar mass range $\rm log(M_\star/M_\odot)= [9.0-11.0]$, and is therefore the dataset of choice in this paper. 

\cite{Belfiore2017a} select star forming galaxies to be moderately face on (major to minor axis ratio greater than 0.4) and exclude interacting and merging galaxies. They calculate metallicity for each star forming region using the \cite{Maiolino2008} metallicity calibration based on the $\mathrm{R_{23} =  ([OII] \lambda\lambda 3726,28+ [OIII]\lambda\lambda 4959, 5007) / H \beta}$ parameter and the \cite{Pettini2004} metallicity calibrator based on $\rm O3N2 = \log ([OIII] \lambda 5007/H\beta) / ([NII] \lambda 6584/H\alpha)$. In this section we fit the \cite{Maiolino2008} abundances, but discuss the differences obtained fitting the \cite{Pettini2004} abundance data in Sec. \ref{sec:3.2}. A discussion on the effect of the metallicity calibration on our results is therefore postponed to that section.

Stacked profiles in mass bins are obtained by computing the robust estimate of the median profile (using Tukey's biweight, \citealt{Beers1990}) and standard deviation within each mass bin. Here we fit the data as a function of physical distance (in kpc) from the galaxy centre. When required, the effective radius is taken to be the median effective radius of the galaxies in each mass bin.

\begin{figure} 
	\includegraphics[width=0.5\textwidth, trim=0 0 0 0, clip]{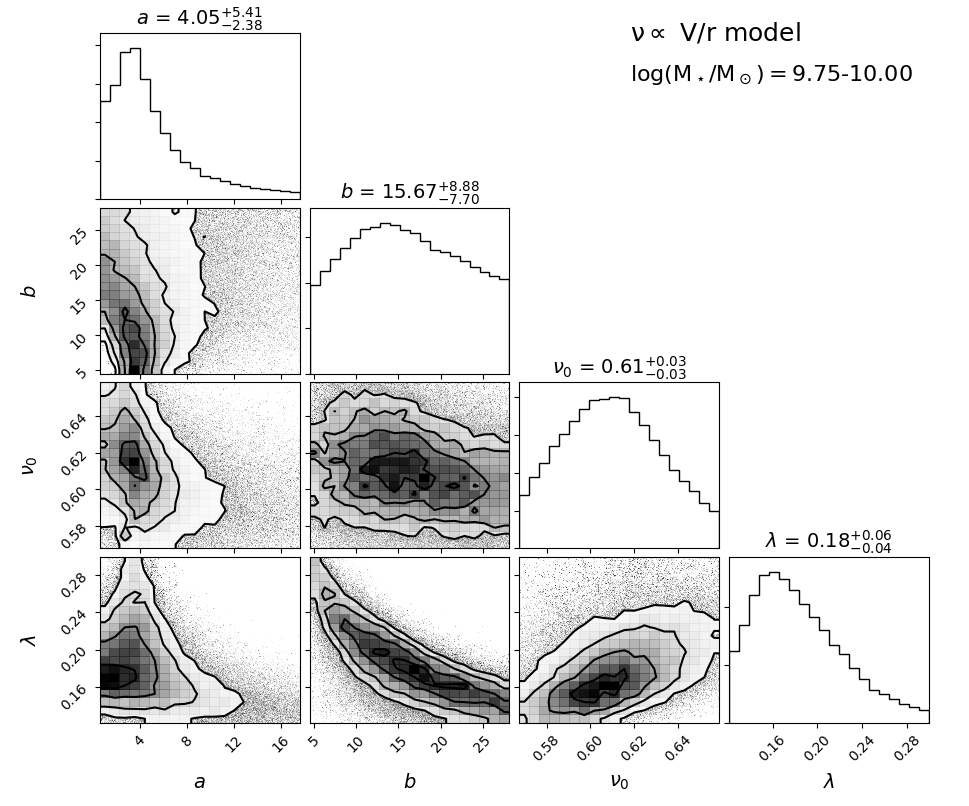}
	
	\caption{Corner plot showing the posterior PDFs of the four model parameters ($a$, $b$, $\nu_0$, and $\lambda$) obtained by fitting the stacked metallicity gradient in the mass bin $\rm log(M_\star/M_\odot)=9.75-10.00$ with the $\rm \nu \propto V/r$ SFE model. The median, 16th and 84th percentiles of the posterior PDF for each parameter are shown above each marginalised PDF.} 
	\label{fig5}
\end{figure}
%

%
\begin{figure*} 
	
	\includegraphics[width=0.70\textwidth, trim=0 0 0 0, clip]{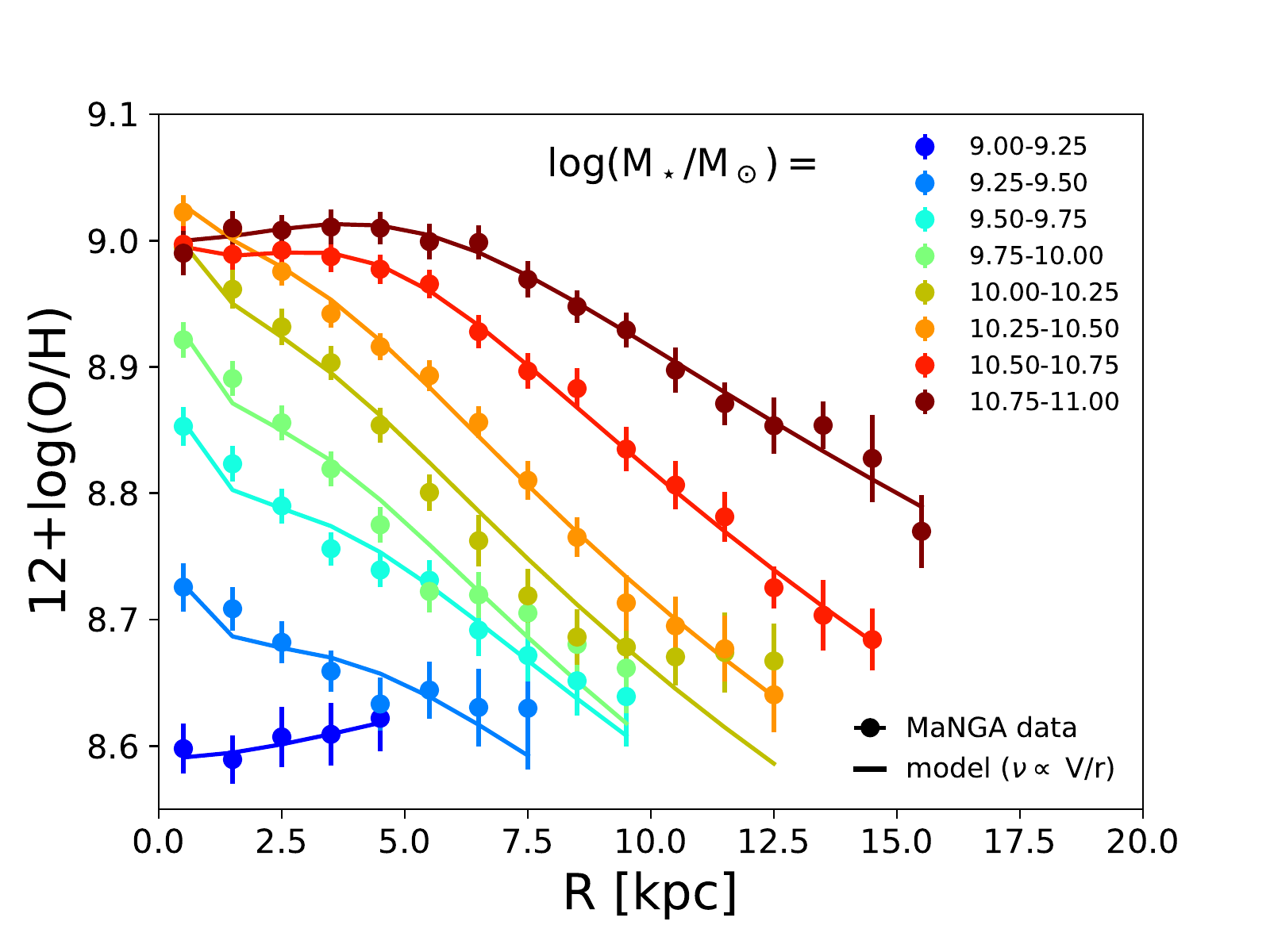}
	\caption{Metallicity gradients in different stellar mass bins from the MaNGA data (\protect\citealt{Belfiore2017a}, coloured circles) and the best-fit chemical evolution model (solid lines). The model used here corresponds to the $\rm \nu \propto V/r$  SFE parametrization. } 
	\label{fig6}
	
	\includegraphics[width=1.\textwidth, trim=0 0 0 0, clip]{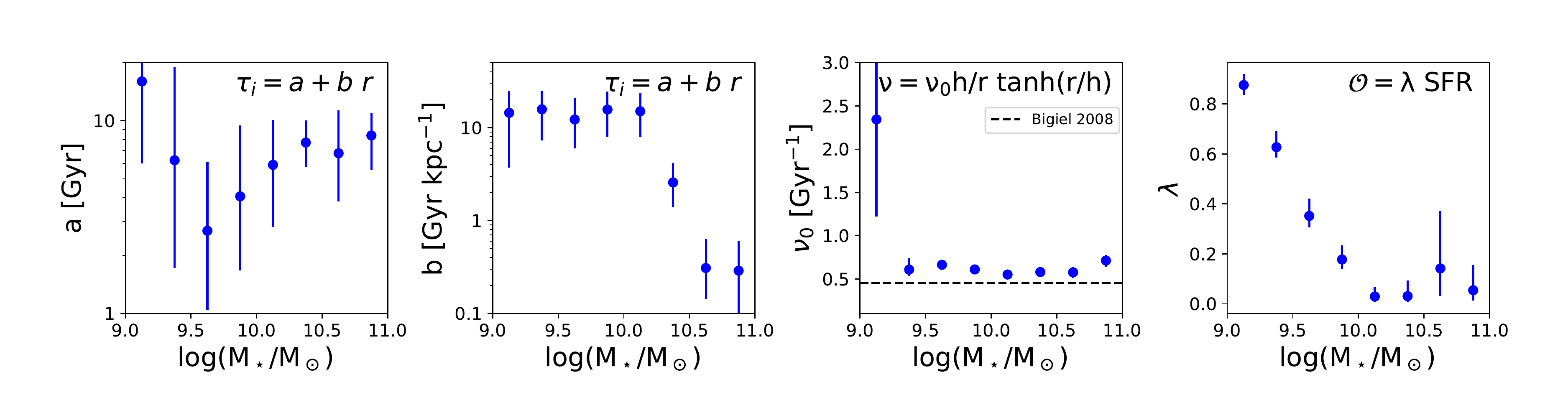}
	\caption{Median values of the PDFs for the model parameters ($a$, $b$, $\nu_0$, and $\lambda$) obtained with the $\rm \nu \propto V/r$  SFE parametrization as a function of stellar mass. The error bars correspond to the 16th and 84th percentiles of the PDF. Also plotted the median $\nu_0$ obtained by \protect\cite{Bigiel2008} for a small sample of local star forming galaxies.} 
	\label{fig7}
\end{figure*}

\begin{figure*} 
	
	\includegraphics[width=0.5\textwidth, trim=0 -50 0 0, clip]{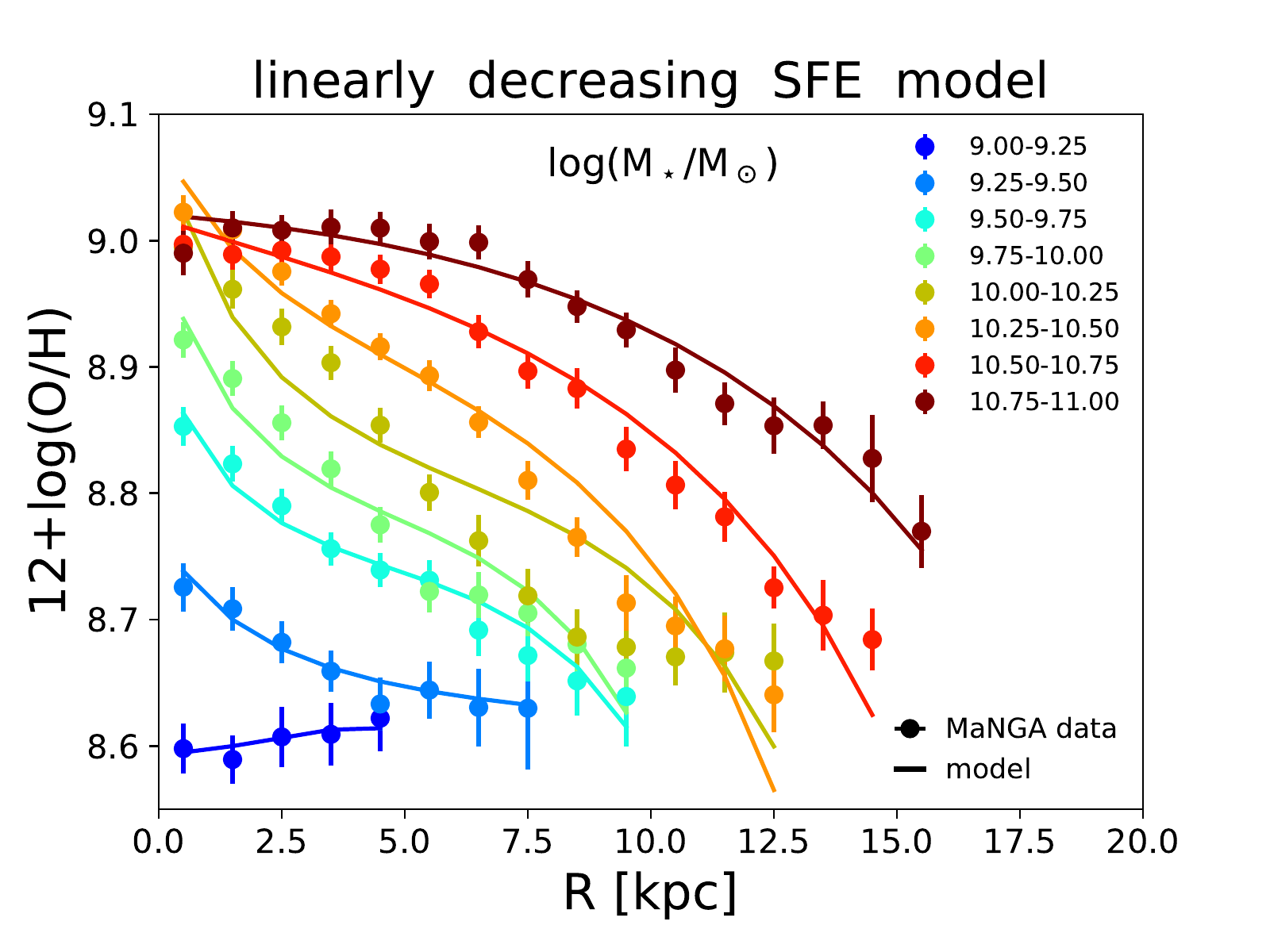}
	\includegraphics[width=0.48\textwidth, trim=0 0 0 0, clip]{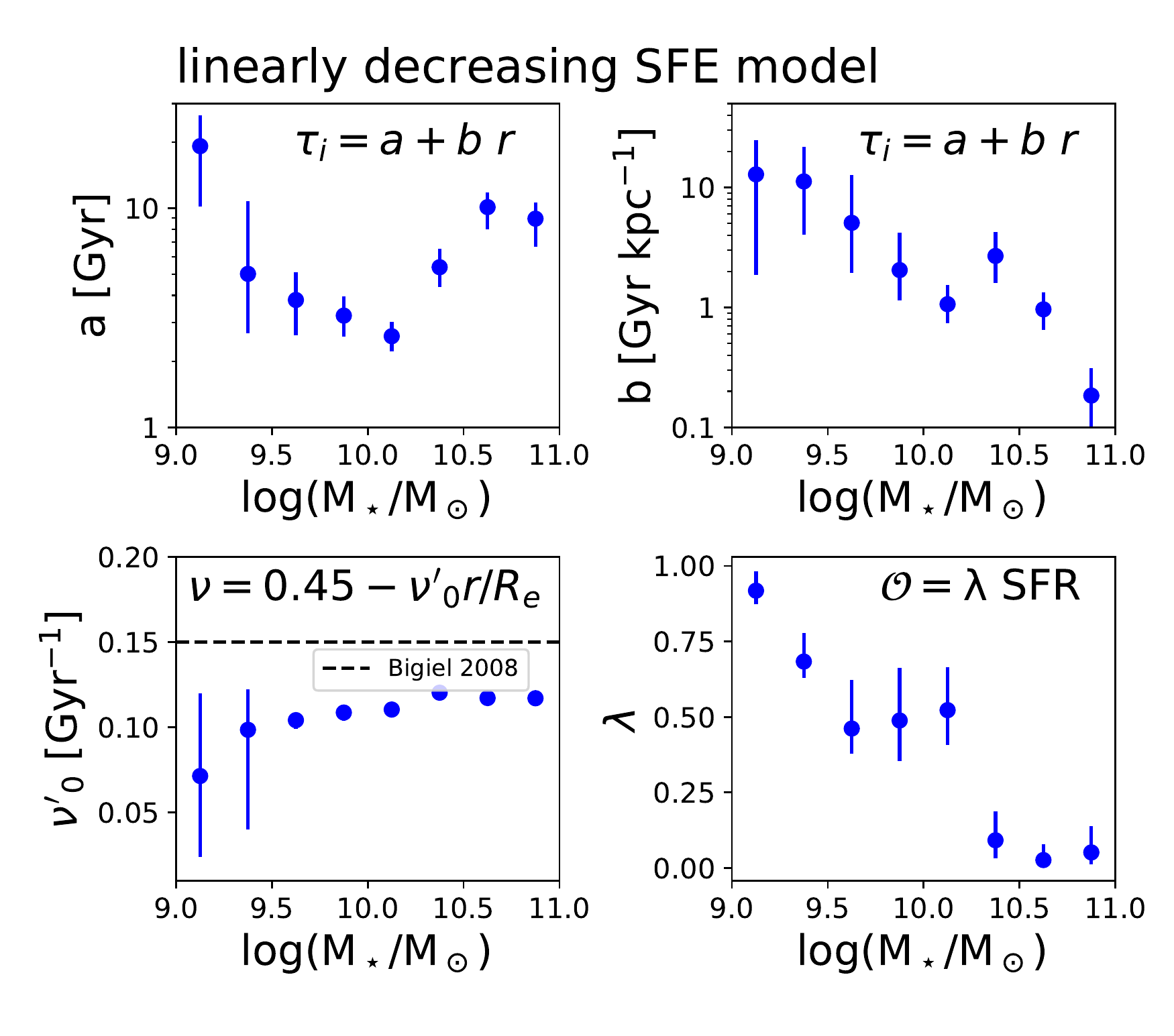}
	\caption{\textit{Left}: Same as \protect\ref{fig6}, but using the linearly decreasing SFE model. \textit{Right}: Same as Fig. \protect\ref{fig7}, but using the linearly decreasing SFE model. The parameter $\nu'_0$ denotes the linear slope of the SFE and is compared to the median value obtained analyzing the data from \protect\cite{Bigiel2008} for a small sample of local star forming galaxies.} 
	\label{fig8}
\end{figure*}

\subsection{The fitting approach}
\label{sec:2:2}

In light of the discussion in Sec. \ref{sec:1.4} we expect significant degeneracies to exist between $a$, $b$, $\nu_0$, and $\lambda$, since these parameters conspire in setting the both normalisation and the slope of the metallicity gradient. In fact, one of the aims of this work is to reveal the degeneracies inherent in bathtub chemical evolution models. To this aim, we make use of an MCMC sampling method to explore the four-dimensional parameter space and efficiently characterize the uncertainties associated with the derived parameters. In detail, we consider a Gaussian likelihood function and assume flat priors for our model parameters in the following ranges: $\rm a = [0, 30]~ Gyr$, $\rm b = [0, 30]~ Gyr/kpc$, $\rm \nu_0= [ 0, 4]~Gyr^{-1}$, $\rm \lambda= [ 0, 10]$. For the model with linearly decreasing SFE, we use a flat prior for $\nu_0'$ in the range [0, 0.3] $\rm Gyr^{-1}$. In all models we assume t $=$ 14 Gyr.

The python module \textsc{emcee} \citep{Foreman-Mackey2013} is used to perform the MCMC sampling using an affine invariant algorithm \citep{Goodman2010}. The sampler is initialized around the position in parameter space which provides a best fit to the data, calculated using a the \textsc{scipy} optimize.minimize procedure.
We fit the metallicity gradients in each mass bin independently, without constraining the model parameter to vary smoothly with mass. The acceptance fractions (i.e. fraction of times a suggested step is approved in the evolution of the Markov chain) is between 0.3 and 0.5 for all the mass bins, indicating reasonable performance for the MCMC sampler. Convergence of the MCMC chain is also evaluated using the Gelman-Rubin diagnostic, $\hat{R}$. This metric compares the dispersion within the chain to the dispersion between chains sampling the same posterior. $\hat{R}$ tends to 1 when convergence is achieved. We find $\hat{R}< 1.08$ for all four free parameters.


In Fig. \ref{fig5} we show the posterior probability density functions (PDF) obtained fitting the metallicity gradient in the mass bin $\rm log(M_\star/M_\odot)=9.75-10.00$ with the $\rm \nu \propto V/r$ SFE model as an example of the kind of degeneracies unveiled by the MCMC analysis. The degeneracy contours appear different for different mass bins, but some general properties are already evident in the example in Fig. \ref{fig5}. In particular, the $a$ and $b$ parameters, determining the gas infall timescale, cannot be inferred very precisely from the data. In this example the PDF for $b$ parameter covers a large fraction of the prior range. The PDFs for $\rm \nu_0$ and $\lambda$, on the other hand, are relatively well-constrained around the best-fit values. 
The most significant degeneracies appear between $b$, $\lambda$ and $\nu_0$. A higher value of $b$ generates a steep gradient at late times, which can also be obtained by a slight decrease of the outflow loading factor, or of $\nu_0$ (see Fig. \ref{fig4}). Corner plots showing the PDFs for the other mass bins and for the linearly decreasing SFE model are shown in Appendix \ref{sec:appB}.

\subsection{The best fit model parameters}
\label{sec:2:3}

Focusing first on the $\rm \nu \propto V/r$ SFE model, we show in Fig. \ref{fig6} the best fit models for each mass bin (solid lines), superimposed on the MaNGA metallicity profiles in mass bins (circles with error bars). It is clear from this figure that the model with $\rm \nu \propto V/r$ provides excellent fits to the data across all mass bins. The reduced $\chi^2$ values for each mass bin lie in the range between 0.3 and 2.4, with an average $\chi^2$ across all mass bins of 1.0.

Our models are capable of matching the change in shape of the metallicity gradients as a function of stellar mass, fitting both the steep gradients in the mass range $\rm log(M_\star/M_\odot)=9.5-10.5$ and the high-mass galaxies, which show a flattening of the metallicity gradient in the central regions. The data for the lowest mass bin shows a mildly inverted gradient, which is also well-fitted by our models. However, the PDFs for $a$, $b$ and $\nu_0$ in the lowest-mass bin are roughly flat over the prior range and only the outflow loading factor $\lambda$ shows a PDF with a well-defined peak for this mass bin (see Fig. \ref{fig:AppB}).
In the mass range $\rm log(M_\star/M_\odot)=9.25-10.00$ a small break is evident in the slope of the model metallicity gradients at $\rm r \sim 2.0$ kpc. This corresponds to the disc scale-length $h$, and to the change in slope of the rotation curve, as parametrized by the $\rm \nu \propto V/r$ model.

In Fig. \ref{fig7} we show the medians of the inferred posterior PDFs for the four model parameters as a function of stellar mass. The error bars represent the 16th and 84th percentiles of the posterior PDFs.
We note two regimes, which roughly correspond to high-mass ($\rm log(M_\star/M_\odot)>10.25$) and low-mass galaxies. At low masses $a$ and $b$ are not well-determined and the outflow loading factor $\lambda$ is a decreasing function of mass. At high masses, roughly corresponding to the onset of the flattening of the metallicity gradient in the central regions, $b$ is low and $\lambda$ is marginally consistent with being zero. Interestingly, $\nu_0$ is very well-constrained in the range 0.5-0.7 $\rm Gyr^{-1}$ across the whole mass range. These values of $\nu_0$ are comparable to the observed value of 0.45 $\rm Gyr^{-1}$ \citep{Bigiel2008}. Considering that we assumed an uninformative prior in the range $\rm [ 0, 4]~Gyr^{-1} $, we consider the fact that inferred $\nu_0$ lies close to the measured value as an additional success of the model.
As noted above, the lowest mass bin is an outlier, and the inferred high median value of $\nu_0$ is entirely due to the flat posterior PDF.

In Fig. \ref{fig8} (left panel) we show the equivalent results for the linearly decreasing SFE model. This model is less successful at reproducing in detail the shape of the MaNGA metallicity radial profiles, with values of the reduced $\chi^2$ going from 0.4 to 7.7 and an average $\chi^2$ over all mass bins of 2.5. The model struggles in particular to reproduce galaxies of intermediate mass $\rm log(M_\star/M_\odot)=10.0-10.5$, where the shape of the metallicity gradient changes from steep to flat in the inner regions. 


The parameters inferred using the linearly decreasing SFE model are shown on the right panel of \ref{fig8}. Several trends are found to be in common with the $\nu \propto V/r$ model. In particular, $\lambda$ shows a decrease with mass, going from $\sim 1$ at $\rm log(M_\star/M_\odot)=9.0$ to zero at high masses. $\nu'_0$, the slope of the SFE decrease in units of $\rm r/R_e$, has only a mild mass dependence and the average value is 0.11 $\rm Gyr^{-1}$, in reasonable agreement with the value of 0.15 $\rm Gyr^{-1}$ derived in Sec. \ref{sec:1.3:2} from the data of \cite{Bigiel2008}.


\begin{figure*} 
	\includegraphics[width=0.75\textwidth, trim=0 0 0 0, clip]{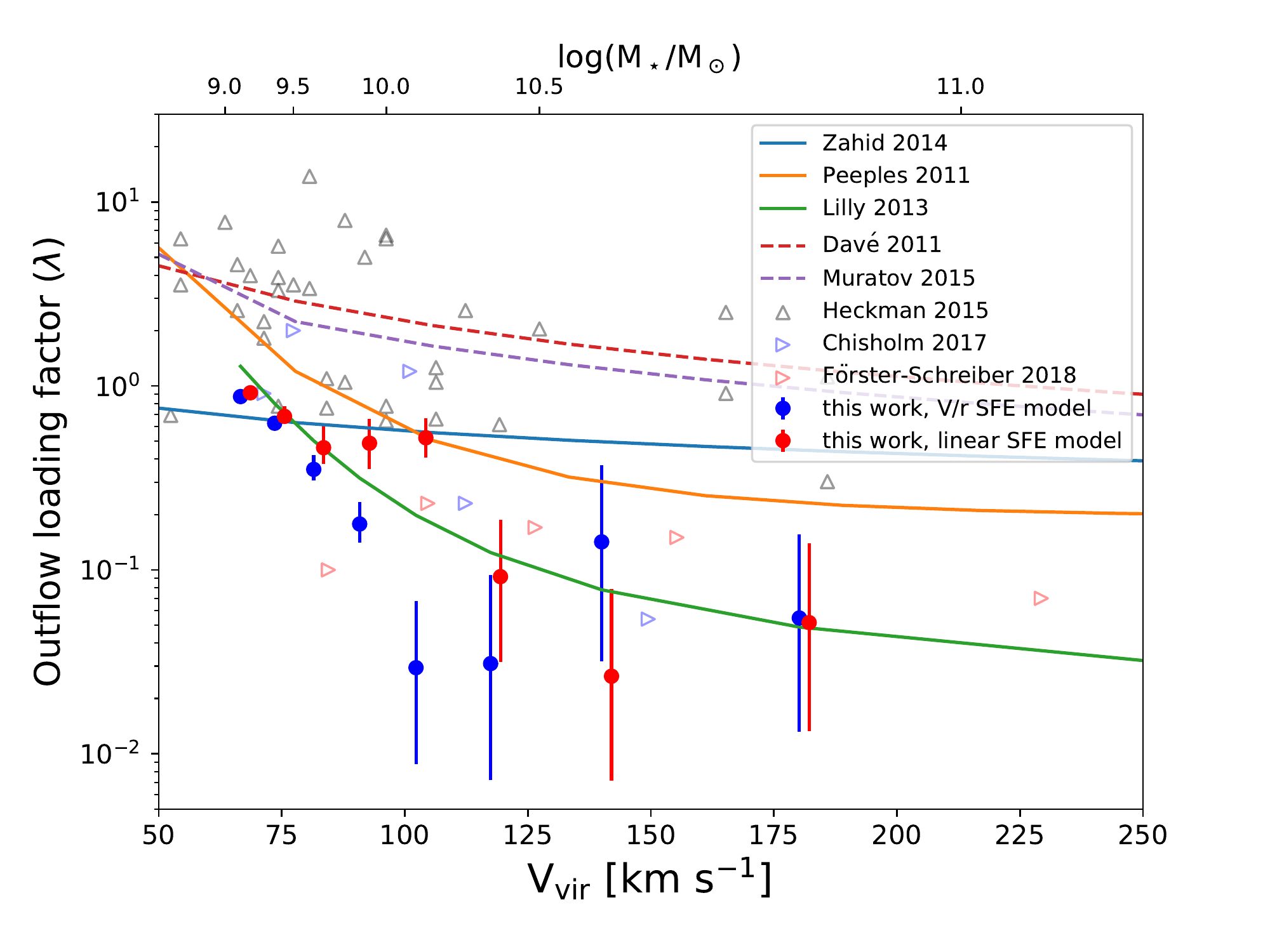}
	\caption{The outflow loading factor as a function of the virial velocity $\rm V_{vir}$ (and stellar mass). The solid blue and red circles correspond to the time averaged cumulative loading factors inferred in this work by fitting the MaNGA metallicity gradients with the $\nu \propto V/r$ and the linearly decreasing SFE model respectively. Solid lines represent empirical determinations of the outflow loading factor by different authors, who fitted the mass-metallicity relation (and sometimes additional information, like gas fraction or SFR). In particular, we show the results from \protect\cite{Peeples2011}, \protect\cite{Lilly2013} and \protect\cite{Zahid2014}. The triangles correspond to instantaneous loading factors estimated from observations of local (upwards black, \protect\citealt{Heckman2015}, right-pointing blue, \protect\citealt{Chisholm2017}) and high-redshift $\rm z =  0.6-2.7 $ galaxies (right-pointing red, \protect\citealt{Forster-Schreiber2018a}). The dashed lines represent mass loading factors measured in the hydrodynamical simulations of \protect\cite{Dave2011} and  \protect\cite{Muratov2015}.}
	\label{fig9}
\end{figure*} 	

\section{Discussion} 
\label{sec:3}

In this work we have demonstrated that our simple chemical evolution models, based on the gas-regulatory formalism and containing only four free parameters, are capable of reproducing the change in \textit{shape} of metallicity radial profiles for galaxies of different stellar masses at z $=$ 0 as observed by the MaNGA survey. Of the two SFE parameterizations considered in this work, the model with $\rm \nu \propto V/r$ produces the best fit to the data. Moreover, the inferred SFE is close to the SFE measured in galaxies in the local Universe by \cite{Bigiel2008}.
While the assumption of inside-out growth is supported by our parameter inference ($b$>0), the overall constraints on the parameters setting the infall timescale ($a$ and $b$) are weak, their mass dependence appears somewhat different between the $\rm \nu \propto V/r$ and the linearly decreasing SFE model. In this section we therefore do not discuss our inference on these parameters. On the other hand, taken at face value, our models are capable of inferring the value of the outflow loading factor with good precision. We therefore focus this discussion on comparing our inferred outflow loading factor with results from the literature and discussing possible systematic uncertainties, such as the value of the nucleosynthetic yield or the gas-phase metallicity calibration used. 

Finally we compare our results with other recent work based on both more sophisticated analytical models and hydrodynamic simulations.

\subsection{The outflow loading factor}
\label{sec:3.1}

In the literature there are at least two different approaches to measuring the outflow loading factor, which can be broadly defined in both observations and simulations. The first approach aims to measure the instantaneous outflows loading factor by relating the state of the outflowing gas directly to an ongoing star formation event. Measurements of outflow rates based on the kinematics of the ionised or molecular gas close to the galactic discs of starburst galaxies fall into this category \citep{Heckman2000, Veilleux2005, Heckman2015, Forster-Schreiber2018a}. Hydrodynamical codes which assume sub-grid prescriptions for launching winds also generally quote the outflow loading factor directly related to a star formation event (also referred to as the loading factor `at injection', see \citealt{Pillepich2018}). The shortcoming of this approach is that this outflow loading factor does not take the fate of the outflowing gas into account, since a large fraction of this gas may be quickly re-integrated into the disc and therefore promptly made available for future star formation. This means that the instantaneous outflow loading factor cannot be directly related to the baryon and metal deficit of galaxies.

Alternatively one can define an average cumulative outflow loading factor as the ratio between the star formation rate and the amount of gas leaving the galaxy's halo (or crossing a surface at a specific distance from the centre of the halo) over a defined timescale \citep{Muratov2015}. This outflow loading factor is directly related to the amount of baryons and metals expelled by a galaxy, and therefore more closely comparable to the outflow loading factor used in chemical evolution models. In presence of recycling, the time averaged cumulative outflow loading factor will be lower than the instantaneous one. In this work one may interpret, albeit approximately, the outflow loading factor computed here as a average of the instantaneous loading factor over the SFH of the galaxy.

Taking these differences into account, we show in Fig. \ref{fig9}  the outflow loading factor inferred in this work as blue and red circles with error bars, corresponding to the $\nu \propto V/r$ and the linearly decreasing SFE model respectively. See Table \ref{table_best_fit_param} for the tabulated values of the loading factors and their errors.
We also show in Fig. \ref{fig9} the outflow loading factors inferred from measurements of the local mass-metallicity relation by authors using different analytical models (solid lines, \citealt{Peeples2011, Lilly2013, Zahid2014}). For consistency with previous literature, the halo mass and virial velocities are obtained from the stellar mass using the formalism and equations of \cite{Peeples2011}. It is worth noting that different authors make use of different metallicity calibrations and oxygen nucleosynthetic yields, which would affect their inference for the outflow loading factor. Considering these systematic uncertainties and the differences in the modeling framework, the outflow loading factors obtained in this work are in reasonable agreement with the range of values present in the literature. 

Coloured triangles in Fig. \ref{fig9} refer to the loading factors inferred `directly' from observations of star forming galaxies at $\rm z \sim 0 $ (black upwards triangles, from \citealt{Heckman2015}, blue right-pointing triangles from \citealt{Chisholm2017}) and $\rm z =  0.6-2.7 $ (red right-pointing triangles, from \citealt{Forster-Schreiber2018a}). These determinations refer to instantaneous loading factors, measured by studying UV and H$\alpha$ line emission respectively. We note that large systematics may affect these determinations, due to uncertainties in the density and geometry of the outflows and because they only refer to a specific phase of the ISM. However, partially because of the large intrinsic scatter, these loading factors are in good general agreement with the loading factors determined in this work.

The dashed lines in Fig. \ref{fig9} show the values for the outflow loading factors employed or measured in hydrodynamical simulations by \cite{Dave2011} and \cite{Muratov2015} for galaxies in dark matter halos of different masses. 
Other simulators only quote loading factors at injection (e.g. \citealt{Vogelsberger2013} and \citealt{Pillepich2018} for the loading factors at injection for the Illustris and Illustris-TNG simulations respectively), which can be up to an order of magnitude larger. We predict however, that since Illustris and TNG reproduce the mass-metallicity relation for galaxies \citep{Vogelsberger2013, Torrey2019}, a direct measurement of cumulative loading factor would be in much better agreement with other measurements of the loading factor in the literature. While other theoretical estimates of the outflow loading factor still lie above most of the empirically-determined values, we will show in the next section that changing the oxygen yield and/or the adopted metallicity calibration one may infer higher loading factors from the observations as well. 

It is also worth noting that relaxing the assumption that outflows share the metallicity of the surrounding ISM will have an impact on the loading factor inferred in this work and other studies based on fitting the mass-metallicity relation. Based on a sample of five galaxies with high signal-to-noise UV spectroscopy, \cite{Chisholm2018} find that for galaxies with $\rm log(M_\star/M_\odot)>9.0$ have outflow metallicities $\sim$ 2.6 time higher than the metallicity of the ISM gas in the host. Unfortunately introducing metal-enriched outflows requires new solution to the constitutive equations of our model, so we do not explore this issue in detail here. 

Finally, we note that the assumption of a loading factor that does not change with time (as adopted in this work) is not a bad one. To test this, we parametrised the instantaneous loading factor as a power law in virial velocity. We then use the SFH of each model galaxy to derive a stellar mass, halo mass and virial velocity as a function of time following the formalism in \cite{Peeples2011}. We find that assuming either $\rm \lambda \propto 1/V_{vir}$ or $\rm \lambda \propto 1/V_{vir}^2$ the SFH-averaged loading factor (calculated at redshift zero) is within 20\% of the instantaneous loading factor.


\begin{figure*} 
	\includegraphics[width=0.45\textwidth, trim=0 0 0 0, clip]{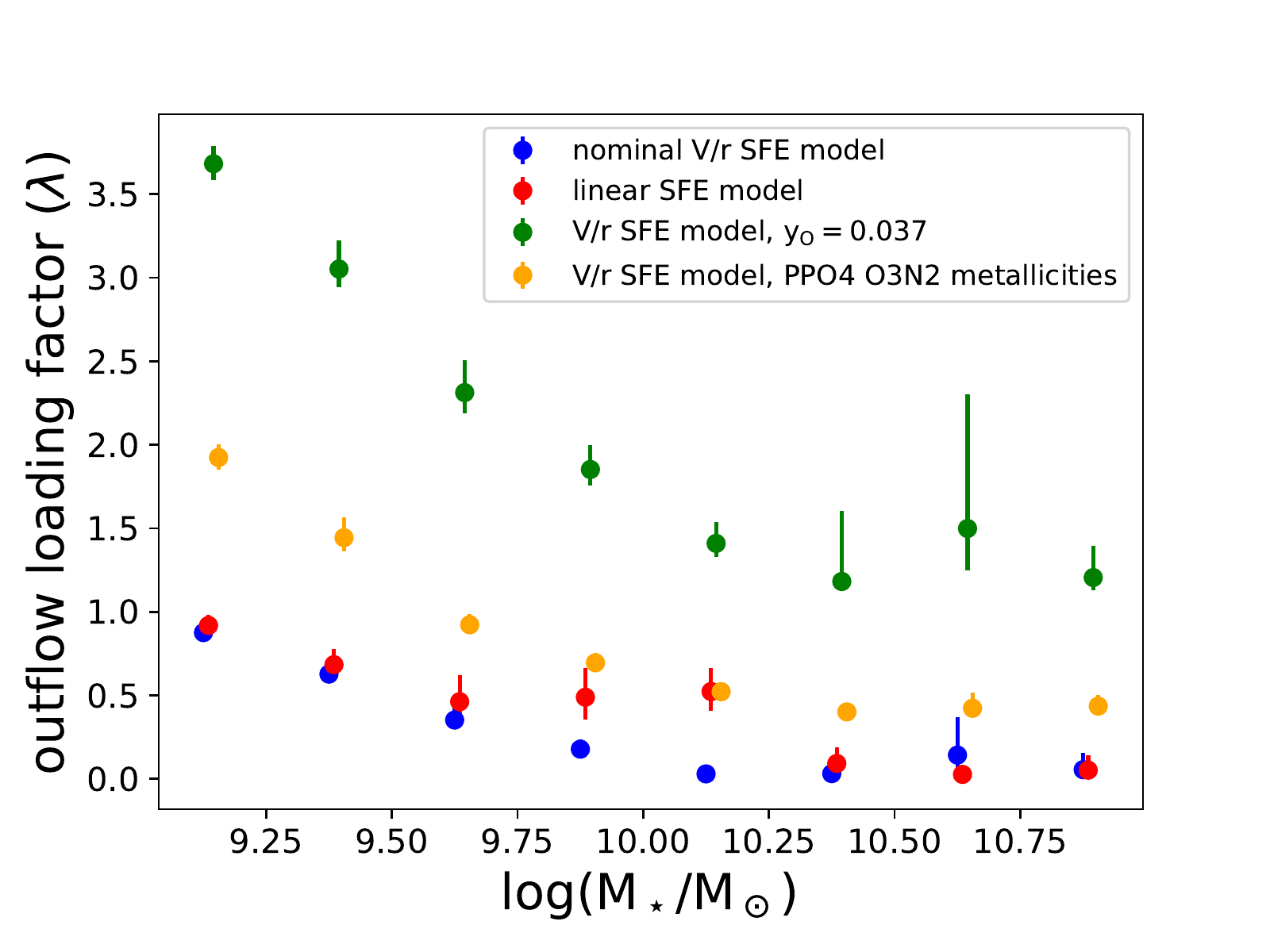}
	\includegraphics[width=0.45\textwidth, trim=0 0 0 0, clip]{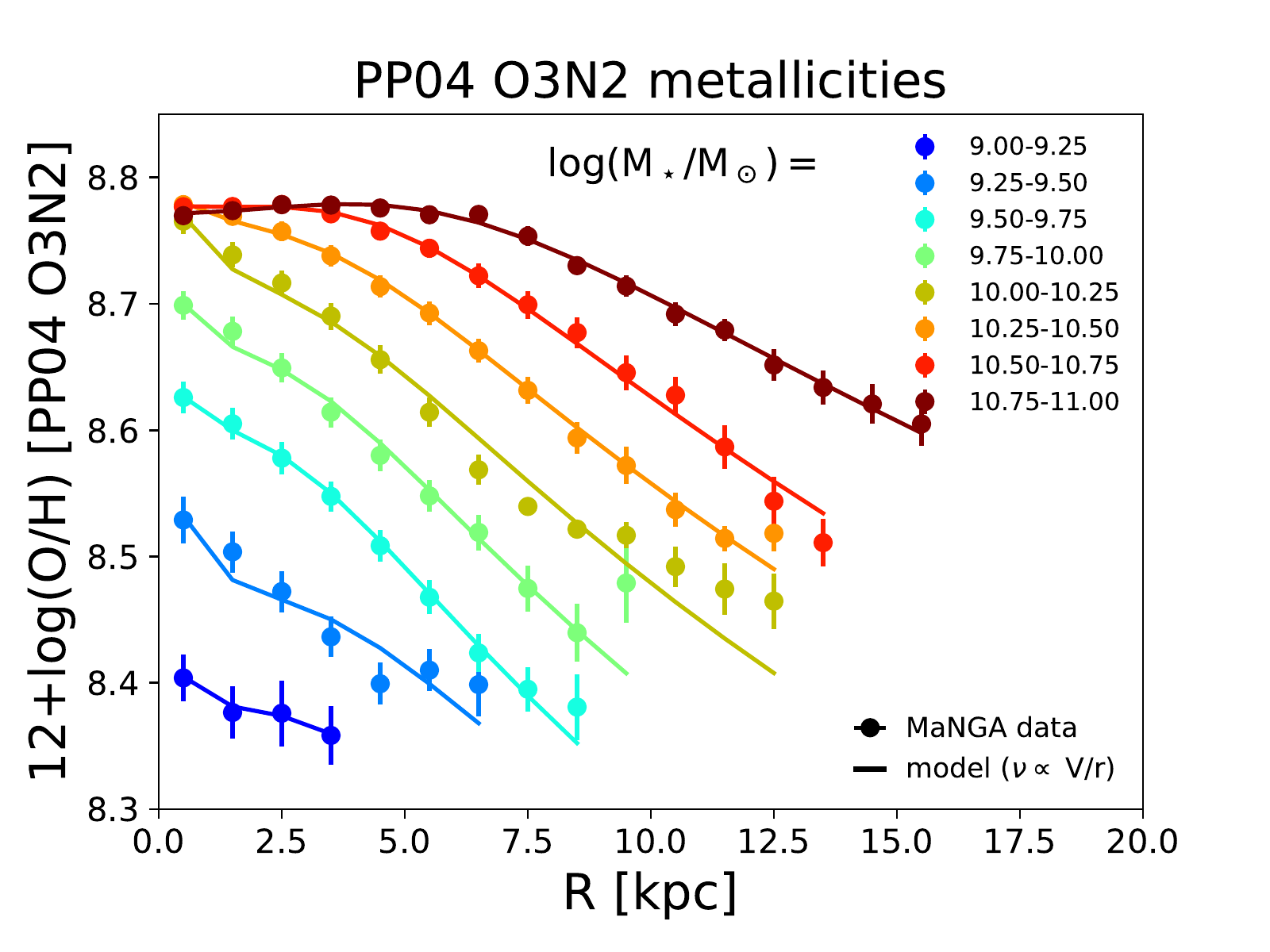}
	\caption{ 
		\textit{Left}: The outflow loading factor as a function of stellar mass computed in this work, using different SFE models ($\nu \propto V/r$ in red, linearly decreasing SFE model in  blue, same as in Fig. \ref{fig9}), different oxygen yield ($\rm y_O=0.037$, as predicted for a \protect\cite{Chabrier2003} IMF, green) and using the \protect\cite{Pettini2004} metallicity calibration based on O3N2 for the metallicity gradients data (orange).
		\textit{Right}: Same as Fig. \ref{fig6}, but using the metallicity gradients derived from the \protect\cite{Pettini2004} calibration based on O3N2.
	}
	\label{fig10}
\end{figure*} 

\subsection{The effect of the yield, IMF and the abundance scale}
\label{sec:3.2}

Both the assumed nucleosynthetic yield and the zero-point of the gas phase abundance scale are expected to have a large effect on the determination of the outflow loading factor. Nearly all oxygen in the Universe is produced by massive stars ($\rm M_\star > 8~M_\odot $) dying as Type II supernovae. Uncertainties in the oxygen yield arise from the systematic uncertainties in predicting the amount of newly synthesized oxygen by Type II supernovae of fixed progenitor mass, but also from the integration over the IMF and the choice of range of stellar masses to integrate over. 

Significant uncertainties persist in modern determinations of the oxygen yield per stellar generation, which can lead to yields varying up to a factor of three for reasonable choices of parameters \citep{Vincenzo2016}. In order to test the impact a higher oxygen yield would have on our analysis we have re-fitted the $\nu \propto V/r$ model to the metallicity profiles data using $\rm y_O= 0.037$, the value expected from a \cite{Chabrier2003} IMF (our default yield is $\rm y_O =0.0105$). The return fraction was also adjusted to $\rm \mathcal{R}=0.455$, as appropriate for this choice of IMF. The resulting outflow loading factor for different mass bin is shown in Fig. \ref{fig10} (green data points), together with the loading factors previously inferred in Sec. \ref{sec:2:3}. As expected, the new yield increases $\lambda$ to values ranging from $\sim 3.5$ for $\rm log(M_\star/M_\odot)=9.0$ to 1.5 at high masses.


Finally we consider the effect of the gas-phase metallicity calibration on the derived loading factor. Well-known systematics affect the measurement of gas-phase metallicity from strong line ratios (see for example the discussion in \citealt{Blanc2015}). These systematics have the largest effect on the determination of the metallicity zero-point, with the \cite{Maiolino2008} calibration based on R23 leading to metallicities $\sim$ 0.2 dex higher than the \cite{Pettini2004} calibration based on O3N2. We therefore repeat our analysis for the $\nu \propto V/r$ model using the metallicity radial profiles calculated using the \cite{Pettini2004} calibration by \cite{Belfiore2017a}. The data and the resulting best-fit models are shown in Fig. \ref{fig10}, right panel. Echoing \cite{Belfiore2017a}, we note that the metallicity gradients calculated using the \cite{Pettini2004} calibration have similar shapes to those calculated using the \cite{Maiolino2008} calibration. Differences include the fact that the lowest mass bin does not show an inverted gradient and the highest mass bin shows a clear plateau, but a less pronounced inversion at small radii. As can be seen from Fig. \ref{fig10}, our models produce excellent fits to the data. The parameter most affected by the change in metallicity calibration is again $\lambda$, although we also find a slightly lower mean $\rm \nu_0$ (0.38 $\rm Gyr^{-1}$) than for the fit to the \cite{Maiolino2008} calibration data.
The resulting outflow loading factor for different mass bins is plotted in Fig. \ref{fig10}, left panel (orange). $\lambda$ now ranges from $\sim 1.9$ at low masses to 0.4 at high masses. 

The loading factors obtained with the different parameter choices discussed in this section are reported for all mass bins in Table \ref{table_best_fit_param}.


\begin{table*}
	\caption{The mass loading factors derived in this work for different models, IMF choices and metallicity calibrations. When not otherwise specified, the oxygen yield is calculated using the \protect\cite{Romano2010} stellar yields and a \protect\cite{Kroupa1993} IMF ($\rm y_O =0.0105$) and we make use of the \protect\cite{Maiolino2008} metallicity calibration based on R23. This information is presented in graphical form in Fig. \protect\ref{fig10}, left panel.}	
	\footnotesize
	\begin{tabular}{ l c c c c}
		\hline
									&  Outflow loading factor &Outflow loading factor &Outflow loading factor &Outflow loading factor \\
		Mass bins 					&  V/r SFE model	 & linearly decreasing & V/r SFE model with &  V/r SFE model\\
		$\rm log(M_\star/M_\odot)=$	& 	 				& SFE model 						& $\rm  y_O$=0.037 & PP04 met. calibration\\
		\hline
		
		9.00$-$9.25  	& $0.88^{+0.04}_{-0.04} $ 		&  $0.92^{+0.06}_{-0.05}   $ 	&  $3.68^{+0.11}_{-0.10}  $ &  $1.92^{+0.08}_{-0.07}  $	\\
		9.25$-$9.50 	& $0.63^{+0.06}_{-0.04}  $ 		&  $0.68^{+0.09}_{-0.05} $ 		&  $3.05^{+0.17}_{-0.11}  $	&  $1.44^{+0.12}_{-0.08}  $\\
		9.50$-$9.75 	& $0.35^{+0.07}_{-0.05}   $ 	&  $0.46^{+0.16}_{-0.08}   $ 	&  $2.31^{+0.19}_{-0.12}  $	&  $0.92^{+0.06}_{-0.05}  $\\
		9.75$-$10.00 	& $ 0.18^{+0.06}_{-0.04}  $ 	&  $0.49^{+0.18}_{-0.13} $ 		&  $1.85^{+0.15}_{-0.10}  $	&  $0.69^{+0.06}_{-0.04}  $\\
		10.00$-$10.25 	& $0.03^{+0.04}_{-0.02}   $ 	&  $ 0.52^{+0.14}_{-0.11} $ 	&  $1.41^{+0.13}_{-0.08}  $	&  $0.52^{+0.05}_{-0.03}  $\\
		10.25$-$10.50 	& $0.03^{+0.06}_{-0.02}  $ 		&  $ 0.09^{+0.10}_{-0.06}  $ 	&  $1.18^{+0.42}_{-0.06}  $	&  $0.40^{+0.03}_{-0.02}  $\\
		10.50$-$10.75 	& $ 0.14^{+0.23}_{-0.11}  $ 	&  $0.03^{+0.05}_{-0.02}   $ 	&  $1.50^{+0.80}_{-0.25}  $	&  $0.42^{+0.09}_{-0.03}  $\\
		10.75$-$11.00 	& $ 0.05^{+0.10}_{-0.04}  $ 	&  $0.05^{+0.09}_{-0.04}   $ 	&  $1.21^{+0.19}_{-0.07}  $	&  $0.43^{+0.07}_{-0.04}  $\\

	\end{tabular}	
	\label{table_best_fit_param}	
\end{table*} 

\subsection{Limitations and comparison with other models}
\label{sec:3.3}

The model described in this work represents an attempt to describe chemical evolution of disc galaxies in the simplest possible terms. Each of the assumptions described in Sec. \ref{sec:1.3} represents a simplification, since important physical processes are neglected.

As already noted in Sec. \ref{sec:1.3:1}, it is difficult to generate an exponential disc following naive disc assembly prescriptions and neglecting radial flows. However, radial flows in discs are a general consequence of angular momentum conservation, since material from the halo accreting onto the disc at a specific radius will not necessarily share the angular moment of the disc at the point of impact \citep{Mayor1981, Lacey1985, Spitoni2011, Pezzulli2016}. Unfortunately the velocities predicted for these radial flows are of a few km/s, and are observationally challenging to distinguish for other non-axisymmetric disturbances in discs \citep{Wong2004, Schmidt2016}. We note, moreover, that to our knowledge there exists no analytical model of disc formation and chemical evolution which naturally produces an exponential disc, even when radial flows are included. Models where the discs are exponential at all time steps (e.g. \citealt{Pezzulli2016, Bilitewski2012}) need to explicitly assume them to be so.

While we successfully fit metallicity gradients, even in the absence of radial flows, the stellar mass profiles of our model galaxies tend to flatten and therefore deviate from an exponential profile within the inner few kpc. This is a natural consequence of inside-out growth in the bath-tub model. The inflow rate in the central regions of galaxies decreases more quickly that in the outskirts, leading to a decrease in SFR and a flattening mass profile. For the same reason, in our models the sSFR always gently increases at large radii. It is interesting to note that using the best-fit model parameters we have derived, the predicted sSFR profiles fail to match in detail the ones observed by the MaNGA survey \citep{Belfiore2018}, especially at small and large radii. The introduction of radial flows, which we delay to future work, may contribute to bringing the model closer to the observations, by providing more gas in the central regions at late times.

The effect of galaxy mergers on metallicity gradients is also neglected in this work. Merging and interacting galaxies are observed to have flatter metallicity gradients \citep{Kewley2010}, as a result of inward gas flows triggered by tidal interactions \citep{Rupke2010, Torrey2012}. \cite{Fu2013} make use of the Munich L-GALAXIES semi-analytical model with a simple prescription for disc disruption during mergers and find that time since the last merger is the quantity that correlates most strongly with the metallicity gradient at redshift zero.

Finally, the procedure for mixing metals into the ISM also has an impact on the derived abundances. In this work we assume that nucleosynthetic products are instantly mixed with the cold ISM in each annulus. However, a fraction of the newly-produced metals may be directly expelled into the hot halo gas \citep{Chisholm2018}, without fist mixing with the cold galaxy ISM, as we have assumed here. Enriched accretion of gas from the halo at late times leads to flatter gradients and a plateau in metallicity in the outer disc. \cite{Bresolin2012}, for example, argue that the metallicity on the outer disc of nearby galaxies flattens to a value of around 0.35 $\rm Z_\odot$. Similar conclusions on the flattening of metallicity gradients at large radii are notably reached by \cite{Sanchez2014}.

The change in slope of the metallicity gradient as a function of mass has not yet been extensively explored in hydrodynamical models. Based on a sample of 32 zoom-in simulations, of which only 9 were evolved to redshift zero, \cite{Ma2016} find a mild steepening of the metallicity gradient with stellar mass, in qualitative agreement with observations at high redshift \citep{Stott2014}. 

More recently, \cite{Tissera2018} compared their predictions from the EAGLE cosmological simulation \citep{Schaye2015} with $\rm z \sim 0$ MaNGA observations of \cite{Belfiore2017a},  demonstrating that EAGLE galaxies have systematically shallower gradients than observed. The EAGLE simulations also shows a large fraction of galaxies with positive metallicity gradients ($\sim$ 40\%), which is not found in observations of the local Universe \citep{Perez-Montero2016}. \cite{Tissera2018} suggest that the overly flat gradients produced in EAGLE could be due to the roughly flat SFE radial profile in the simulated galaxies, at odds with current observations. It is also possible, as argued in \cite{Ma2016}, that the `effective feedback' model implemented in the EAGLE simulation may artificially mix metals on large scales, thus preventing strong metallicity gradients from forming. The development of more physically-motivated models for feedback and ISM physics may therefore be needed in order to reproduce the changes in slope of the metallicity gradients observed by MaNGA.

\section{Conclusions} 
\label{sec:4}

In this work we have developed analytical chemical evolution models, based on the bathtub model formalism, to describe radial metallicity profiles in local galaxies. We bridge the gap between previous bathtub models, mostly aimed at describing the chemical evolution of galactic systems as a whole (e.g. \citealt{Bouche2010, Lilly2013}), and classical chemical evolution models developed for the Milky Way galaxy (e.g. \citealt{Chiappini2001}). In particular, we adopt the inside-out growth formalism of \cite{Matteucci1989}, which posits a radially-dependent infall timescale, and develop two models for the radial dependence of the SFE. In one version of the model we assume that the SFE is inversely proportional to the orbital timescale (the $\nu \propto V/r$ model), and in the second one we assume a constant SFE for the molecular gas (i.e. $\rm SFR/M_{H2}=const$) and a molecular gas fraction which decreases with radius (the linearly decreasing SFE model), motivated by the data from \cite{Bigiel2008}.

For either SFE parametrization, our models are described by four free parameters. Two of the parameters describe the infall timescale and its radial dependence ($a$ and $b$). For the $\nu \propto V/r$ model, the SFE at the centre of the galaxy is taken as a free parameter, while for the linearly decreasing SFE mode the slope of the radial gradient of the SFE is assumed to be free. The final free parameter is the outflow loading factor.  

We have studied the effect of varying these parameter on the metallicity gradient and its time evolution. Overall, our models predict a \textit{flattening} of the metallicity radial profile with time, in general agreement with results from hydrodynamical simulations and classical chemical evolution models with a radially decreasing SFE. However, all four parameters conspire to set the final degree of chemical enrichment and the slope of the metallicity gradient at late times, pointing to significant degeneracies.

We compare our models with the metallicity radial profiles measured by \cite{Belfiore2017a} for star forming galaxies in the MaNGA survey. We perform the fit within a Bayesian framework and explore the parameter space via MCMC sampling. We summarise the main conclusions from this analysis below.

\begin{enumerate}
	\item {Both SFE parametrisations produce good fits to the data, with the $\nu \propto V/r$ model being favoured in terms of $\chi^2$. Notably, our models are capable of reproducing the details of the changes in shape of the metallicity radial profiles over the entire mass range $\log(M_\star/M_\odot)$=[9.0-11.0] covered by the observations.}
	\item{ We find significant degeneracies between model parameters. Partly as a consequence of that, the inference for the parameters describing the infall rate and its radial dependence is weak.}
	\item{ For both the $\nu \propto V/r$ and the linearly decreasing SFE model we find the best-fit parameters describing the SFE have only a weak mass dependence and are in reasonable agreement with the observations of \cite{Bigiel2008}.}
	\item{For the adopted value of the nucleosynthetic yield ($\rm y_O =0.0105$, assuming a \citealt{Kroupa1993} IMF), the outflow loading factor is found to vary from nearly unity at $\rm \log(M_\star/M_\odot)$ = 9.0 to close to zero at $\rm \log(M_\star/M_\odot)$ = 11.0. These loading factors are in good agreement with previous determinations of the loading factor based on the mass-metallicity relation of local galaxies and with `direct' measurements of the loading factors of local and high-redshift star forming galaxies.}
	\item{ A higher value of the yield ($\rm y_O= 0.037$, as expected from a \citealt{Chabrier2003} IMF) leads to higher inferred loading factors, going from $\sim$ 3.5 for $\rm \log(M_\star/M_\odot) =9$  to close to $\sim$ 1.5 for $\rm \log(M_\star/M_\odot) = 11.0$ .
	The choice of the gas-phase metallicity calibration also has an effect on the inference on the outflow loading factor. Higher loading factors are obtained making use of the \cite{Pettini2004} O3N2 metallicity calibration instead of the adopted \cite{Maiolino2008} calibration based on R23.}
\end{enumerate}

Although our models are successful at reproducing the data and provide physical insight into the effect of different parameters, they do not include all the physics relevant to metallicity gradients. We expect that next generation of  hydrodynamical simulations will be able to study the changes in shape of the metallicity gradients as a function of mass and quantify the impact of the physics which is missing from our simplified framework.

%
%
%
%
%
\newpage
\appendix 
 
\section{Solutions to notable chemical evolution models and the role equilibrium}
\label{sec:appA}

In this Appendix we aim to develop a physical understanding for the solutions of the chemical evolution model presented in this work. We start from notable solutions of simple models and comment on how the solution to the model presented in this work are related to them, especially at late times when the systems tends to equilibrium.



In the presence of inflows and outflows, the simplest gas regulatory models are `equilibrium' models, where the gas mass of the system is taken to be constant in time ($\rm d\Sigma_{g} /dt = 0$, \citealt{Finlator2008, Genel2008, Bouche2010, Genel2012, Dave2012}). Assuming $\rm d\Sigma_{g} /dt = 0$ Eq. \ref{eq1} reduces to 
\begin{equation}
\rm \mathcal{I}  = (1-\mathcal{R}+\lambda) \ \Sigma_{SFR}  \quad \text{ (in equilibrium)}. \label{eqaa1}
\end{equation}
Under these assumption at late times equation \ref{eqz} reduces to
\begin{equation}
Z  = \frac{p}{1-\mathcal{R}+\lambda} \quad \text{ (in equilibrium)}. \label{eqaa_2}
\end{equation}

\begin{figure} 
	\includegraphics[width=0.5\textwidth, trim=0 0 0 0, clip]{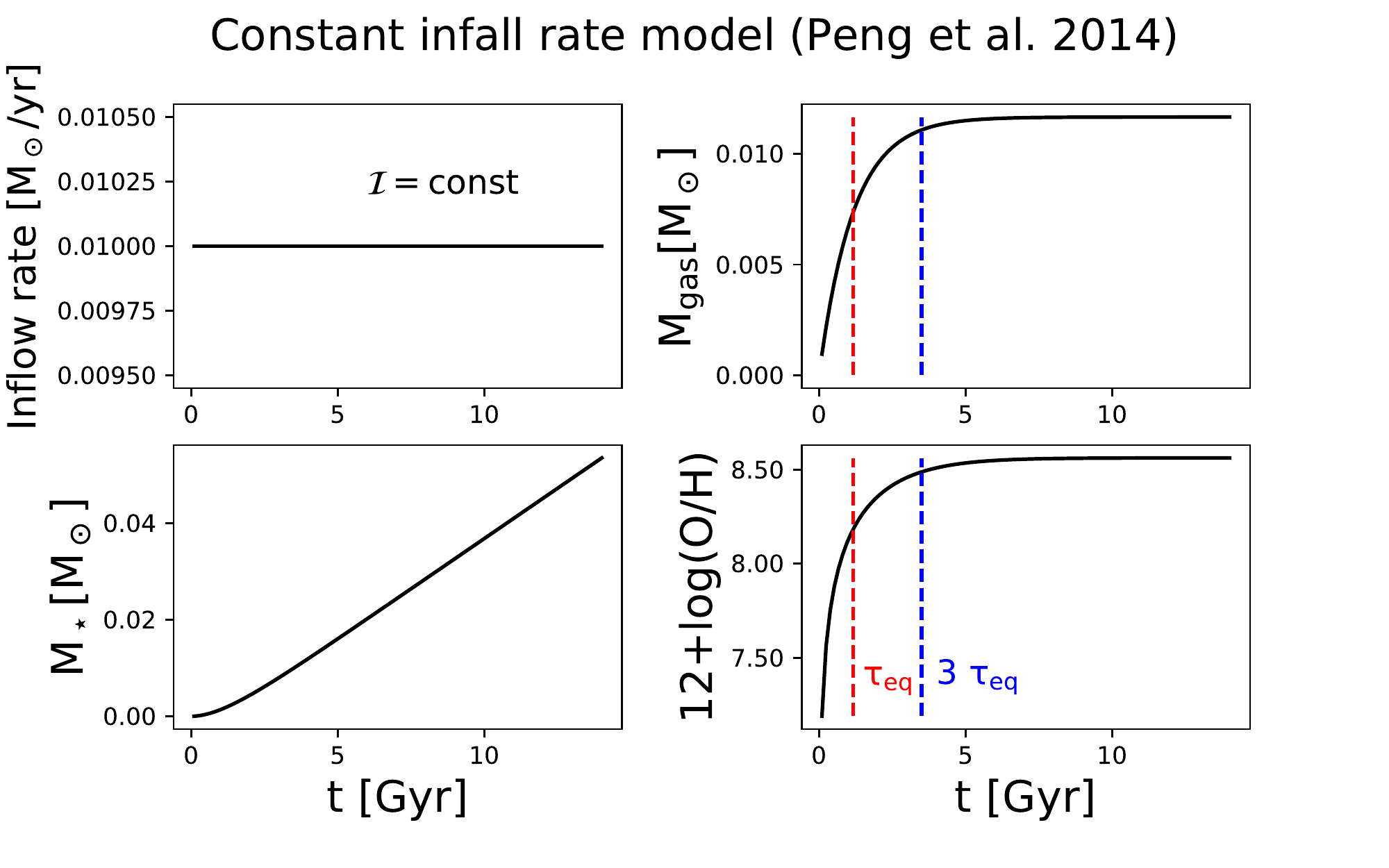}
	\caption{ The time evolution of several fundamental parameters ($\mathcal{I}, \rm M_{gas}, M_\star, 12+log(O/H)$) in the \protect\cite{Peng2014} chemical evolution model with constant infall rate. The metallicity converges towards its equilibrium value at $\rm t > \tau_{eq} = 1.16 ~ Gyr$ (red dashed lines) for the choice of parameters adopted in this plot, $\rm (\nu, \lambda) = (0.5 \ Gyr^{-1}, 1.0)$. Also shown in the plot as blue dashed line the time $\rm t/ \tau_{eq}=3$. We find that for $\rm t /\tau_{eq}>3$ model parameters are sufficiently close to their equilibrium values to justify the equilibrium assumption. } 
	\label{figAPP0}
\end{figure}

\cite{Lilly2013} have highlighted that equilibrium is a good assumption for gas-poor, chemically evolved, low-redshift galaxies, but not for gas-rich dwarfs or galaxies at high redshift. In these systems the SFR cannot adjust itself fast enough compared to the high rate of accretion and the amount of mass in the gas reservoir must evolve. 

If the timescale over which the accretion rate changes is much longer than other timescales in the system, one may assume the inflow rate to be constant to first order. Under these assumptions, \cite{Peng2014} demonstrated that one can define a natural timescale for the chemical evolution of the system (the equilibrium timescale, $\rm \tau_{eq}$), given by

\begin{equation}
\rm \tau_{eq} \equiv \frac{1}{\nu~(1-\mathcal{R}+\lambda)}. \label{eq_tau_eq}
\end{equation}  
For the sake of completeness, we provide below a brief recap on how to obtain analytical solution to the constitutive equations of the bathtub model in the case of constant accretion rate. The evolution of the gas content (Eq. \ref{eq2}) becomes
\begin{equation}
\rm \frac{d \Sigma_g}{dt} + \frac{\Sigma_g}{\tau_{eq}} = \mathcal{I}.
\end{equation}
This is a first order ordinary differential equation with integrating factor $\rm e^{t/\tau_{eq} } $ and solution 
\begin{equation}
\rm \Sigma_g = \mathcal{I} \tau_{eq} ( 1 - e^{- t / \tau_{eq} } ).
\end{equation}
Substituting into equation \ref{eqz} we obtain a first order ordinary differential equation for the time evolution of metallicity 
\begin{equation}
\rm \frac{dZ}{dt} + \frac{Z}{\tau_{eq} (1 - e^{-t/\tau_{eq}}) }  = p ~ \nu,
\end{equation}
which may be solved with integrating factor $\rm e^{t / \tau_{eq} } -1 $. The final time evolution of the metallicity can therefore be written as
\begin{equation}
\rm Z= p~ \nu~ \tau_{eq} - \frac{p ~\nu~ t ~\ e^{-t/\tau_{eq}}  }  {1-e^{-t/\tau_{eq}}}. \label{eq_metall_sol1}
\end{equation}
This solution was already presented in \cite{Belfiore2016}. Confusingly, it is slightly different from the solution originally derived by  \cite{Peng2014}, who assume that the gas mass is slowly varying in deriving their equation 35, while our solution does not make this assumption.

In Fig. \ref{figAPP0} we show the time evolution for a number of fundamental parameters ($\mathcal{I}, \rm M_{gas}, M_\star, 12+log(O/H)$) in this model. The gas phase metallicity of the system increases quickly at early times, and for $\rm t>>\tau_{eq}$ the system tends to the equilibrium metallicity given by
	\begin{equation}
	\rm Z= p~ \nu~ \tau_{eq} \quad \text{ (in equilibrium)} . 
	\label{eq_metall_sol3}
	\end{equation}	
The value of $\rm \tau_{eq}$ is noted in Fig. \ref{figAPP0} as a red dashed line. In practice, in order to test whether the equilibrium condition ($\rm t >> \tau_{eq}$) is met we find that the value of  $\rm t/\tau_{eq}=3$ provides a reasonable boundary (blue dashed line in figure). For example, for the choice of parameters adopted in Fig. \ref{figAPP0} the metallicity is within 80\% of its equilibrium value by $\rm t/\tau_{eq}=2.7$. The gas content, on the other hand, reaches its equilibrium value on a slightly faster timescale. For example, in this model the gas content is within 80\% of its equilibrium value by $\rm t/\tau_{eq}=1.6$.


In order to obtain a time-dependent solution for the equations of chemical evolution using the exponential infall prescription, as we have done in this work, it is useful to define a new timescale, $\rm \tau_c$, given by
\begin{equation}
\rm \tau_{c}^{-1} \equiv \tau_{eq}^{-1}- \tau_{inf}^{-1}.  \label{tc}
\end{equation}
Finding solutions to the constitutive equations of the bathtub model then proceeds in a similar way as for the constant infall rate model.

In particular, in the exponential infall model the time evolution of gas phase metallicity is given by
\begin{equation}
\rm Z= p~ \nu~ \tau_{c} - \frac{p ~\nu~ t ~\ e^{-t/\tau_{c}}  }  {1-e^{-t/\tau_{c}}}. \label{eq_metall_sol2}
\end{equation}
Remarkably this equation is the same as \ref{eq_metall_sol1}, if one substitutes $\rm \tau_{eq}$ with the new timescale $\tau_{c}$. $\tau_{c}$ can therefore be thought as the natural timescale for the exponential infall model to reach the equilibrium metallicity. In this model, at late times the metallicity also converges to its equilibrium value, now given by
\begin{equation}
\rm Z= p~ \nu~ \tau_{c} \quad \text{ (in equilibrium)}. \label{eq_metall_eq}
\end{equation}
Solutions for the time evolution of other quantities are summarised in Table \ref{table_sol} of Sec. \ref{sec:1.2}.

Let us now consider the best fit parameters obtained from fitting metallicity gradients with the $\rm \nu \propto V/r$ model. In this model $\rm \tau_{eq}$ increases with radius, since it is inversely proportional to $\rm \nu$. We find $\rm \tau_{eq} \sim 1-2$ Gyr in the centres of galaxies, increasing to 4 $-$ 8 Gyr at $\rm R= 2 ~R_e$ (excluding the lowest mass bin, which has anomalously high $\nu$, resulting in low $\rm \tau_{eq}$). $\tau_c$ follows a similar radial gradient, with some changes due to the effect of $\rm \tau_{inf}$. Given the definition of $\rm \tau_c$,  $\rm \tau_{inf}$ has the largest effect on $\rm \tau_c$ when $\rm \tau_{inf} \sim \tau_{eq}$. In the inner regions of galaxies  $\rm \tau_{inf} $ is a few Gyr, comparable to $\rm \tau_{eq}$, and therefore  $\rm \tau_c$ is appreciable larger than $\rm \tau_{eq}$. 

\begin{figure} 
	\includegraphics[width=0.5\textwidth, trim=0 0 0 0, clip]{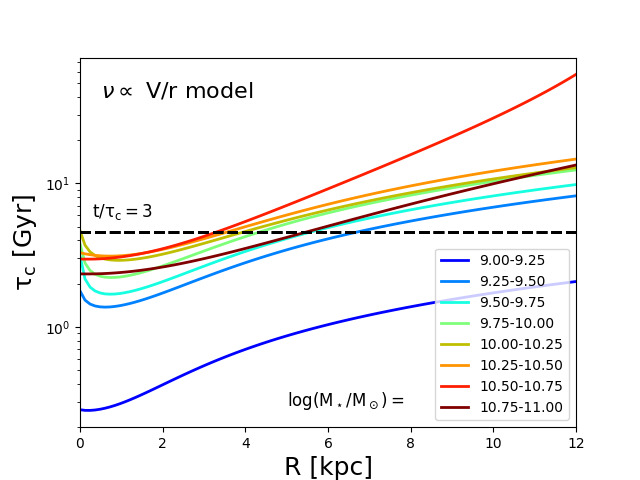}
	\caption{ The radial dependence of $\rm \tau_c$ using the best-fit parameters for the $\nu \propto V/r$ model fit to the metallicity gradients (see Sec. \ref{sec:2:3}). Different colours corresponds to different mass bins, as described in the legend. The dashed black dashed lines corresponds to $\rm t/\tau_c = 3$ and can be used as a rough demarcation between regions in equilibrium (below the line) and regions out of equilibrium (above the line).} 
	\label{figAPP1}
\end{figure}

In Fig. \ref{figAPP1} we show the radial variation of $\rm \tau_c$ using the best-fit model parameters from Sec. \ref{sec:2:3} using the $\nu \propto V/r$ model. Each color represents a different mass bin. In order to give a rough idea of the regions where the galaxies are close to equilibrium, the dashed black line represents $\rm t/\tau_c =3.0$. To first order, regions lying below this line can be assumed to be in equilibrium, while regions lying above this line have not yet reached equilibrium. As can be seen from Fig. \ref{figAPP1}, the inner regions of galaxies are predicted to be in equilibrium (at t $=$ 14 Gyr), while the outskirts increasingly deviate from equilibrium conditions. This is generally true for all mass bins, expect the lowest mass galaxies, as already noted above.

\begin{figure} 
	\includegraphics[width=0.5\textwidth, trim=0 0 0 0, clip]{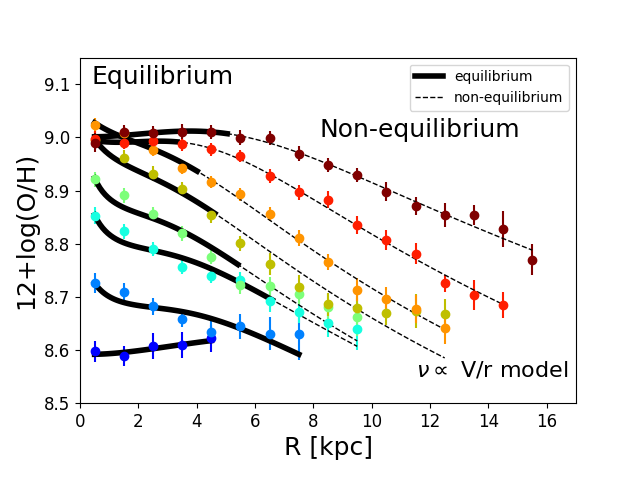}
	\caption{Same as Fig. \ref{fig6}, but dividing highlighting whether each radial regions is the model is or not in equilibrium (solid black and dashed black lines respectively). For the purposes of this plot regions are defined to be in equilibrium if their metallicity is within 0.1 dex of the equilibrium metallicity (Eq. \ref{eq_metall_eq}). } 
	\label{figAPP2}
\end{figure}

In Fig. \ref{figAPP2} we show the best-fit metallicity gradient models (same as Fig. \ref{fig6}), but distinguishing between regions in equilibrium (solid black lines) and regions outside equilibrium (dashed black lines). Here we consider regions to be in equilibrium if the difference between the equilibrium metallicity (Eq. \ref{eq_metall_sol2}) and the metallicity at t$=$14 Gyr is less than 0.1 dex. Again we find that the inner regions of galaxies have already reached their equilibrium metallicity at t$=$14 Gyr and the outer regions are still not in equilibrium.

This analysis demonstrates, therefore, that the ability to capture the non-equilibrium evolution of the system is key in reproducing the shapes of the metallicity gradients in our modelling framework. An equilibrium model may, on the other hand, be successful at reproducing the abundances in the central regions of galaxies.

\section{Corner plots for all mass bins}
\label{sec:appB}

\begin{figure*} 
	\includegraphics[width=0.37\textwidth, trim=0 0 0 0, clip]{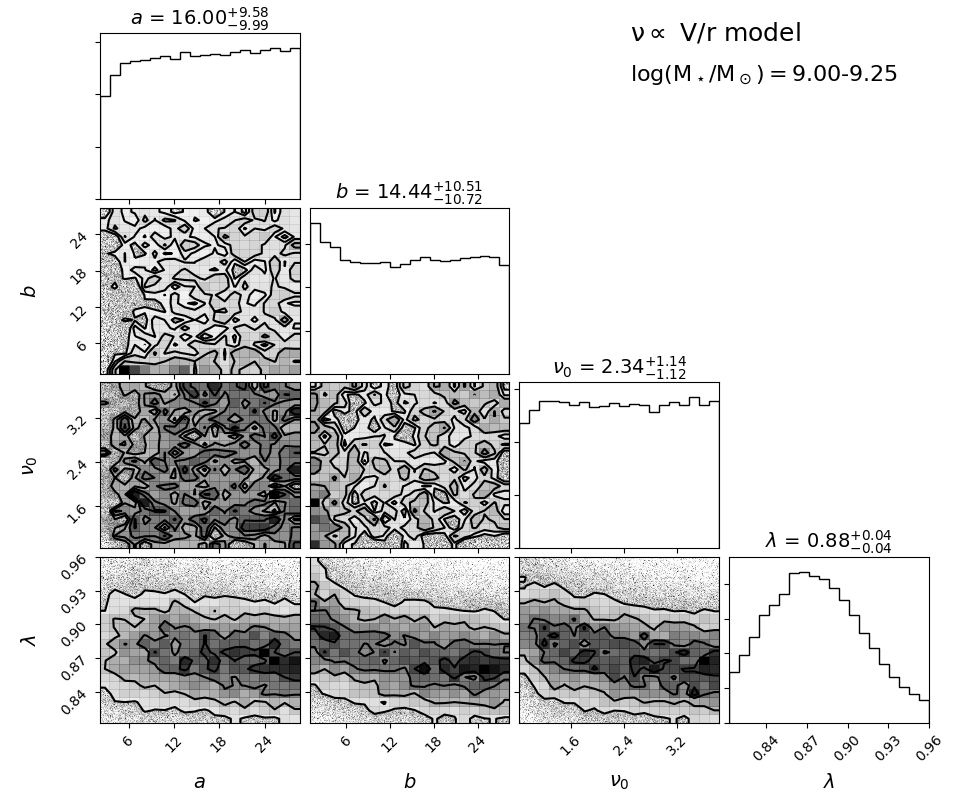}
	\includegraphics[width=0.37\textwidth, trim=0 0 0 0, clip]{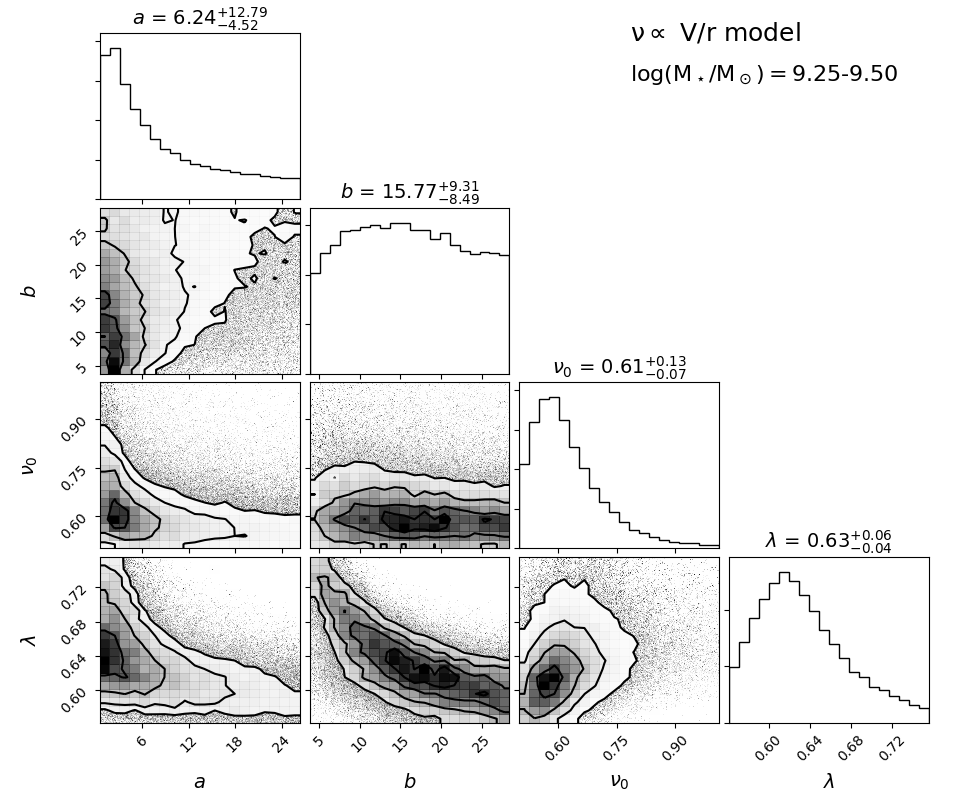}
	\includegraphics[width=0.37\textwidth, trim=0 0 0 0, clip]{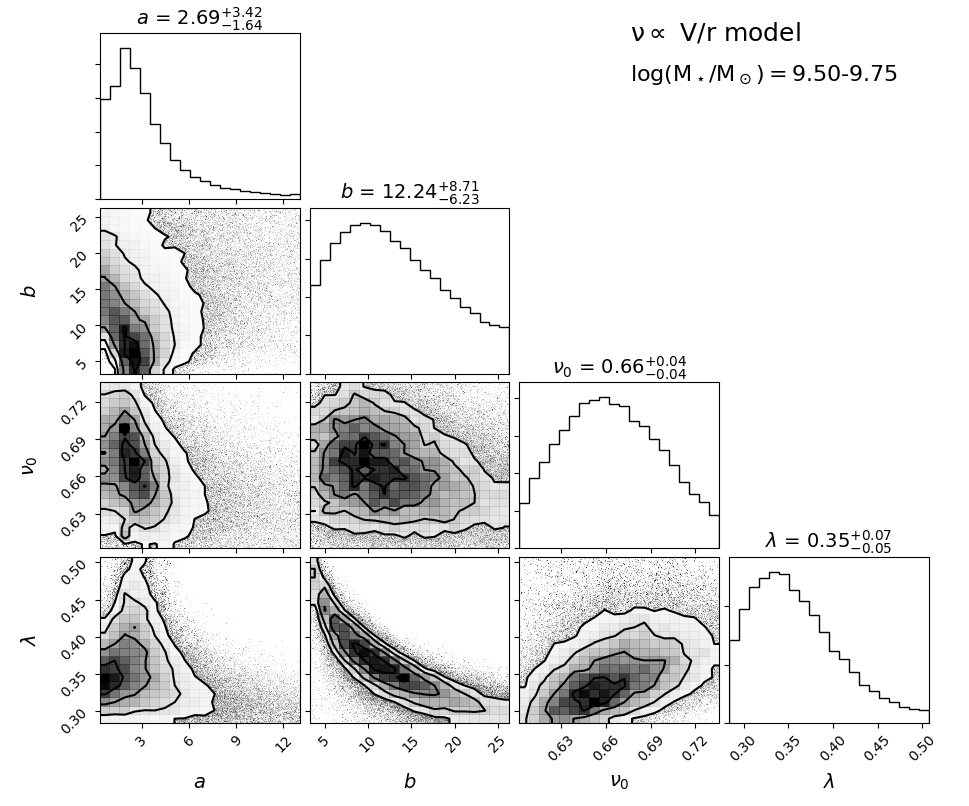}
	\includegraphics[width=0.37\textwidth, trim=0 0 0 0, clip]{fig/pretty_corner3_apr2018_20walkers_20000chains_5000burn.png}
	\includegraphics[width=0.37\textwidth, trim=0 0 0 0, clip]{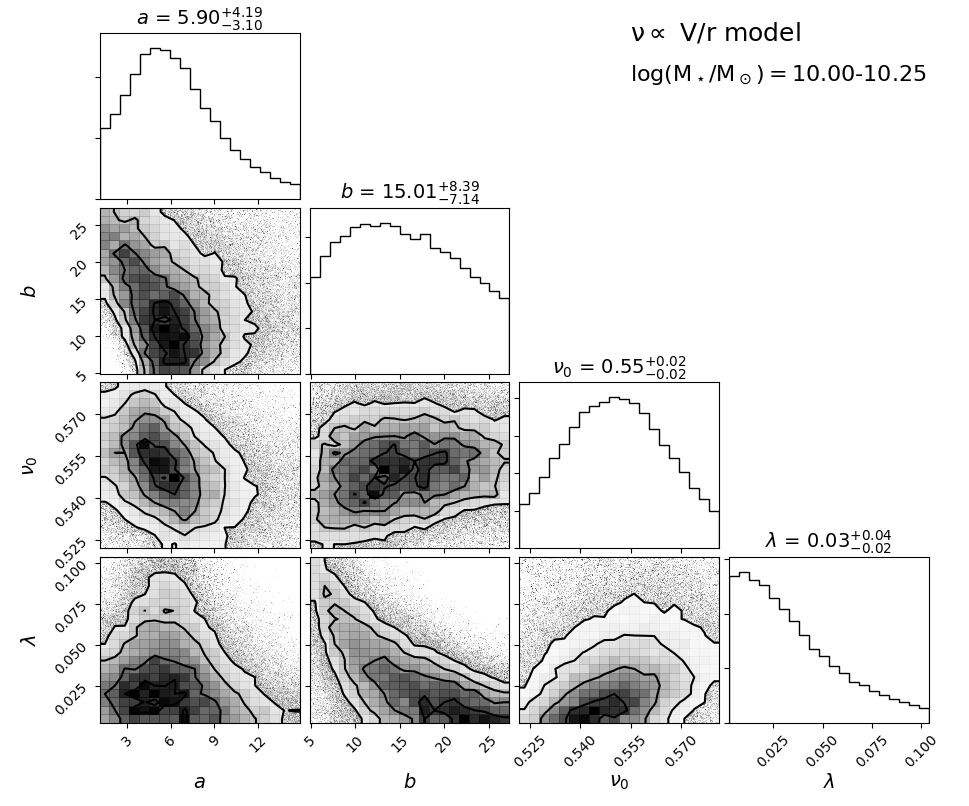}
	\includegraphics[width=0.37\textwidth, trim=0 0 0 0, clip]{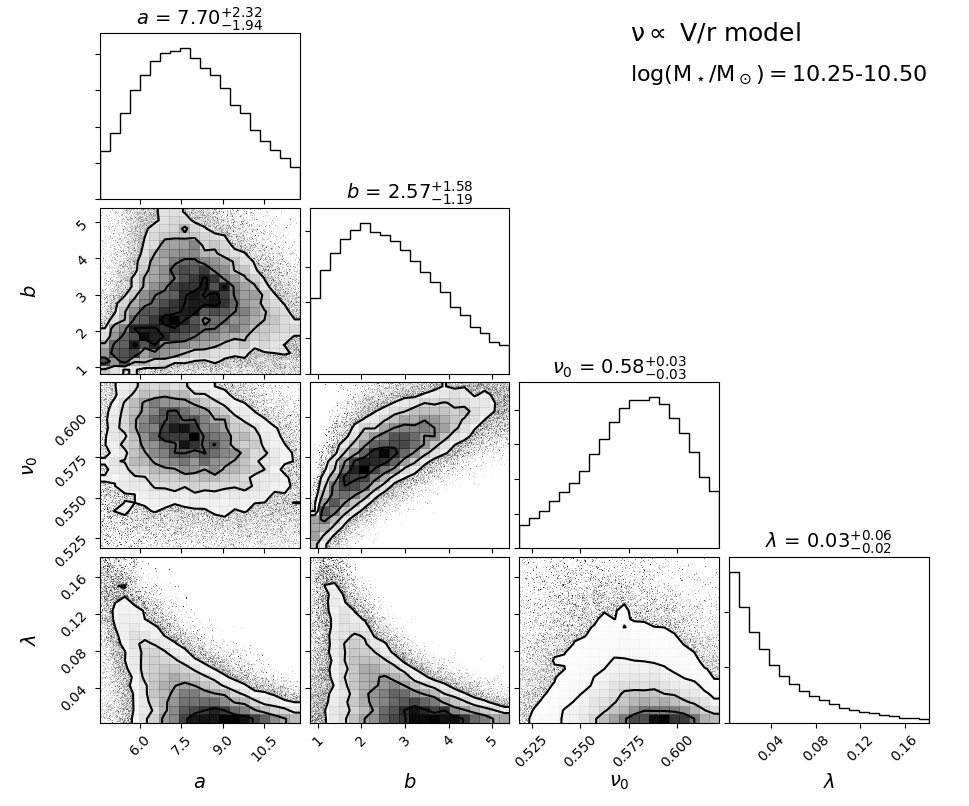}
	\includegraphics[width=0.37\textwidth, trim=0 0 0 0, clip]{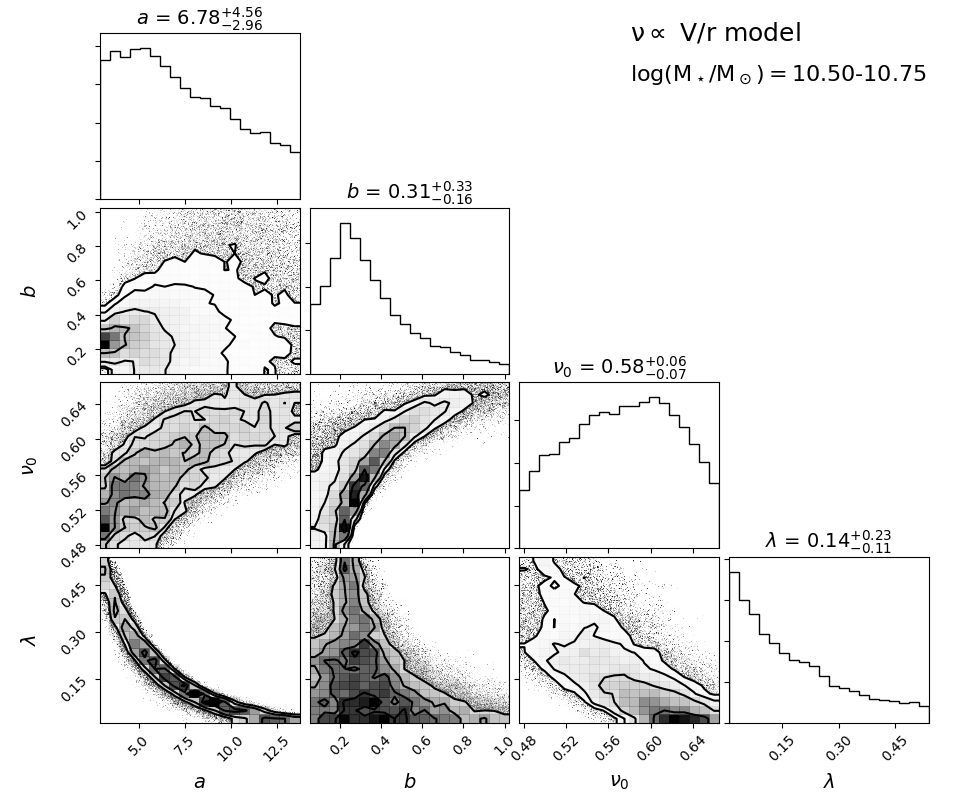}
	\includegraphics[width=0.37\textwidth, trim=0 0 0 0, clip]{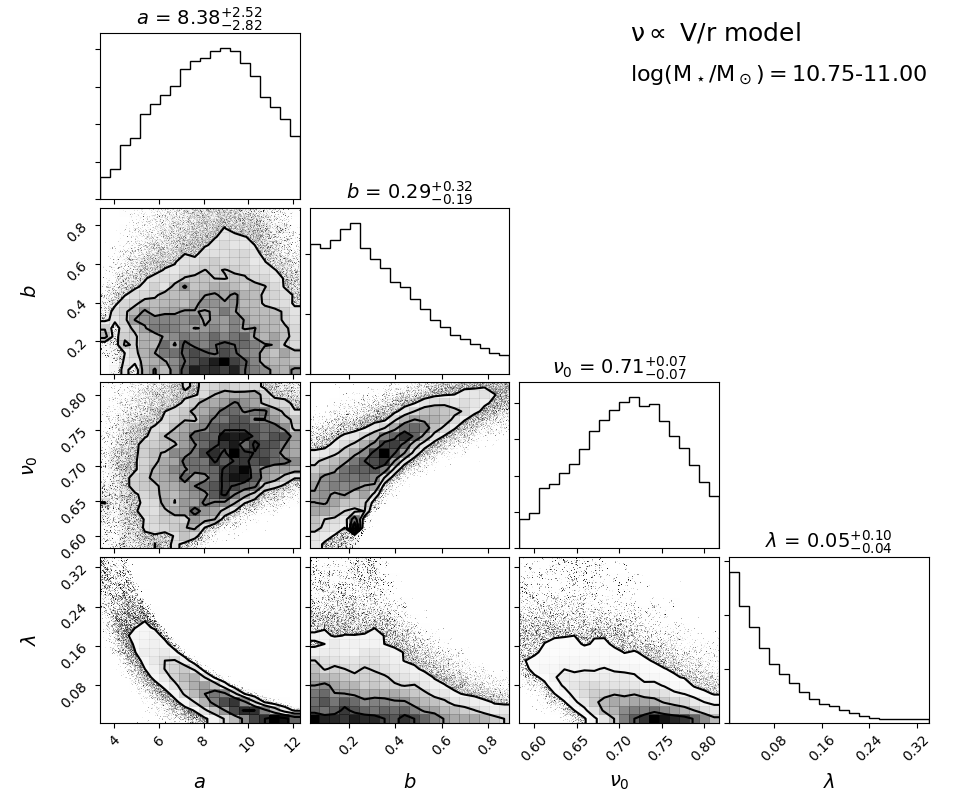}
	
	\caption{Corner plot showing the posterior PDFs of the four model parameters ($a$, $b$, $\nu_0$, and $\lambda$) for the stacked metallicity gradient for all mass bins, fitted with the $\rm \nu \propto V/r$ SFE model. The median, 16th and 84th percentiles of the posterior PDF for each parameter are shown above each marginalised PDF.} 
	\label{fig:AppB}
\end{figure*}

\begin{figure*} 
	\includegraphics[width=0.37\textwidth, trim=0 0 0 0, clip]{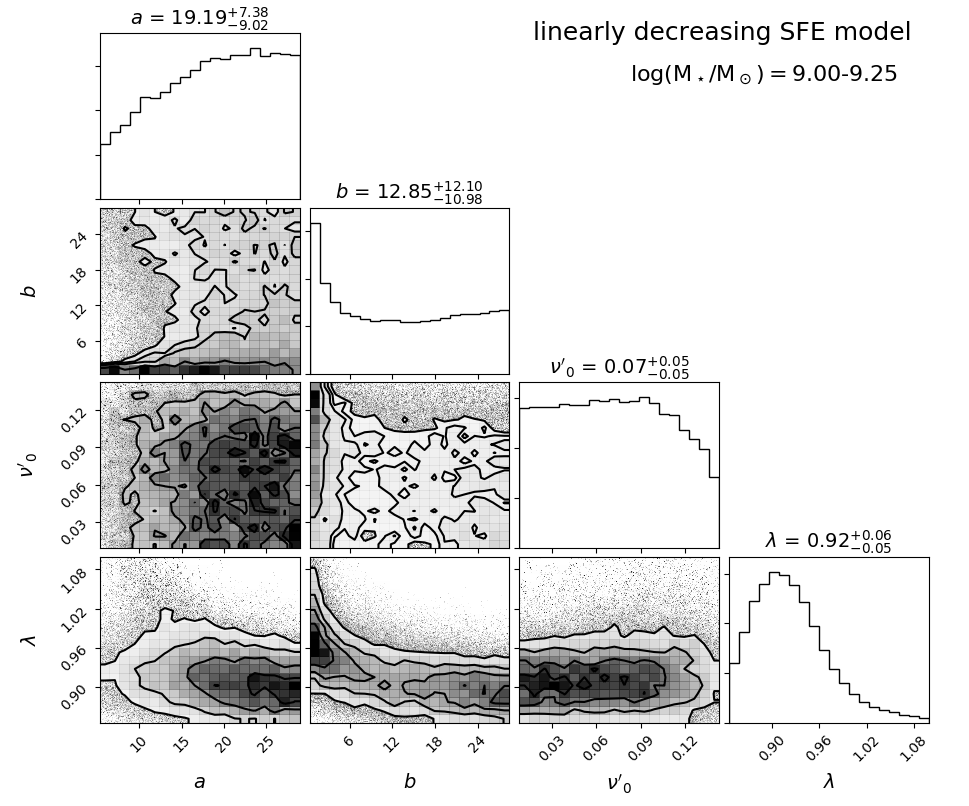}
	\includegraphics[width=0.37\textwidth, trim=0 0 0 0, clip]{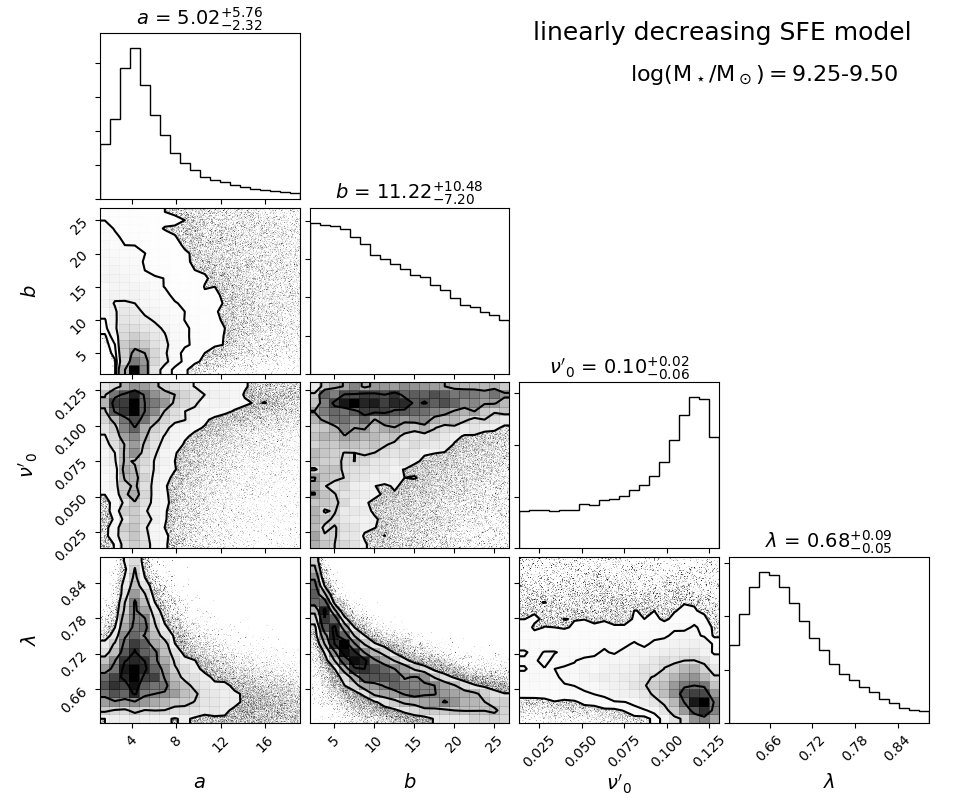}
	\includegraphics[width=0.37\textwidth, trim=0 0 0 0, clip]{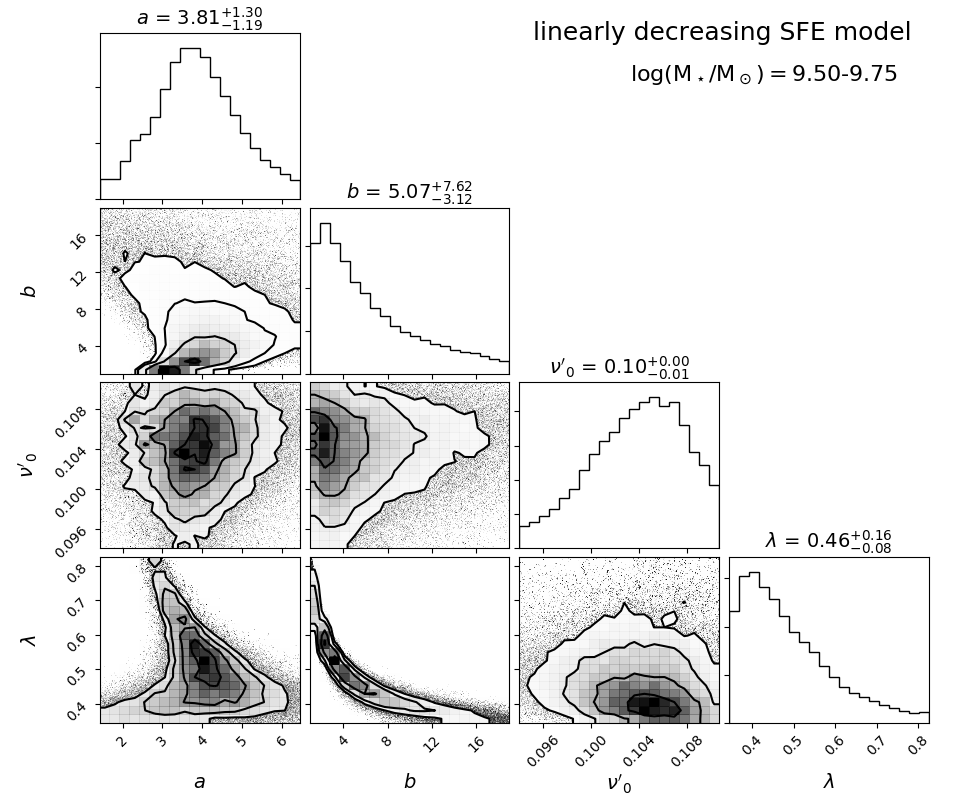}
	\includegraphics[width=0.37\textwidth, trim=0 0 0 0, clip]{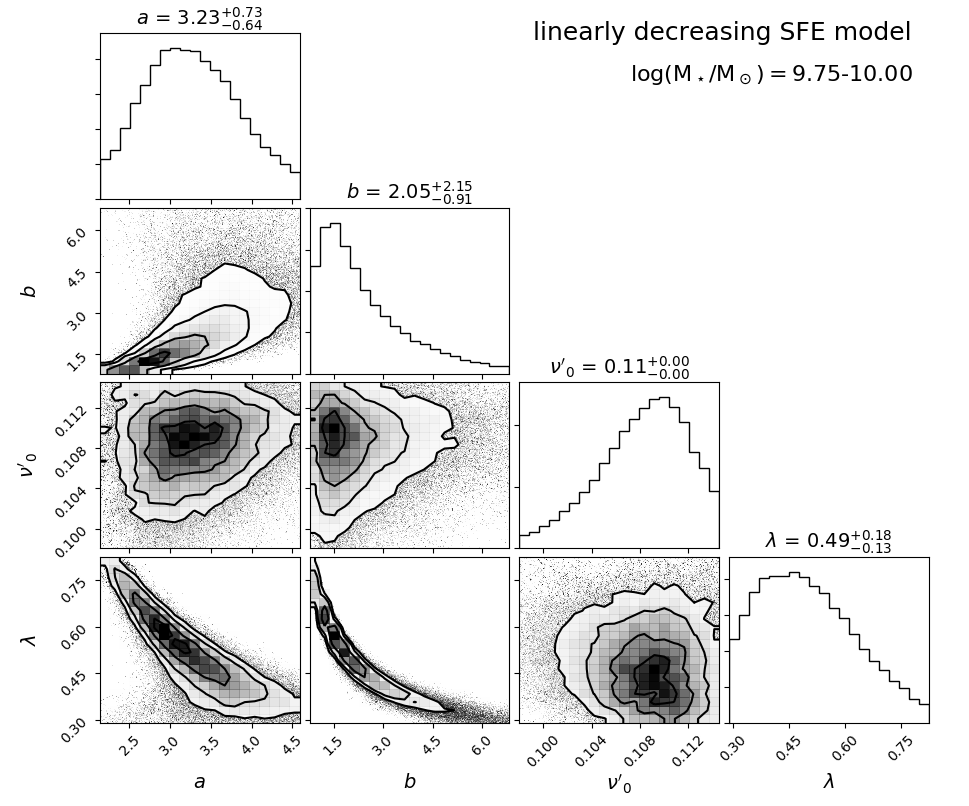}
	\includegraphics[width=0.37\textwidth, trim=0 0 0 0, clip]{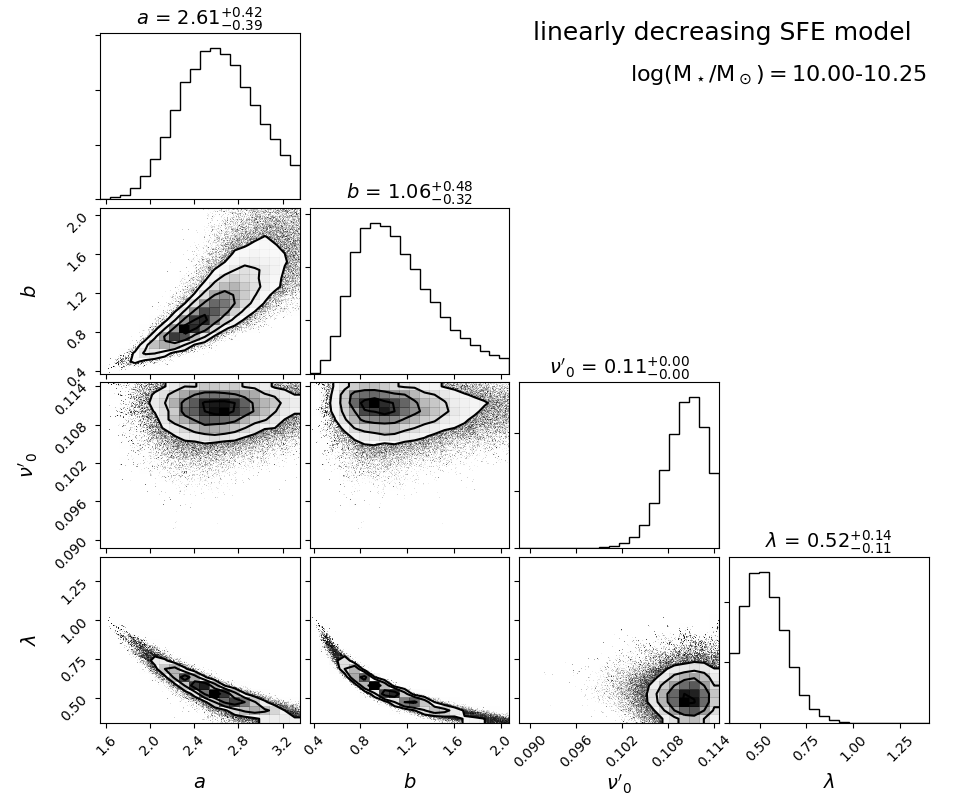}
	\includegraphics[width=0.37\textwidth, trim=0 0 0 0, clip]{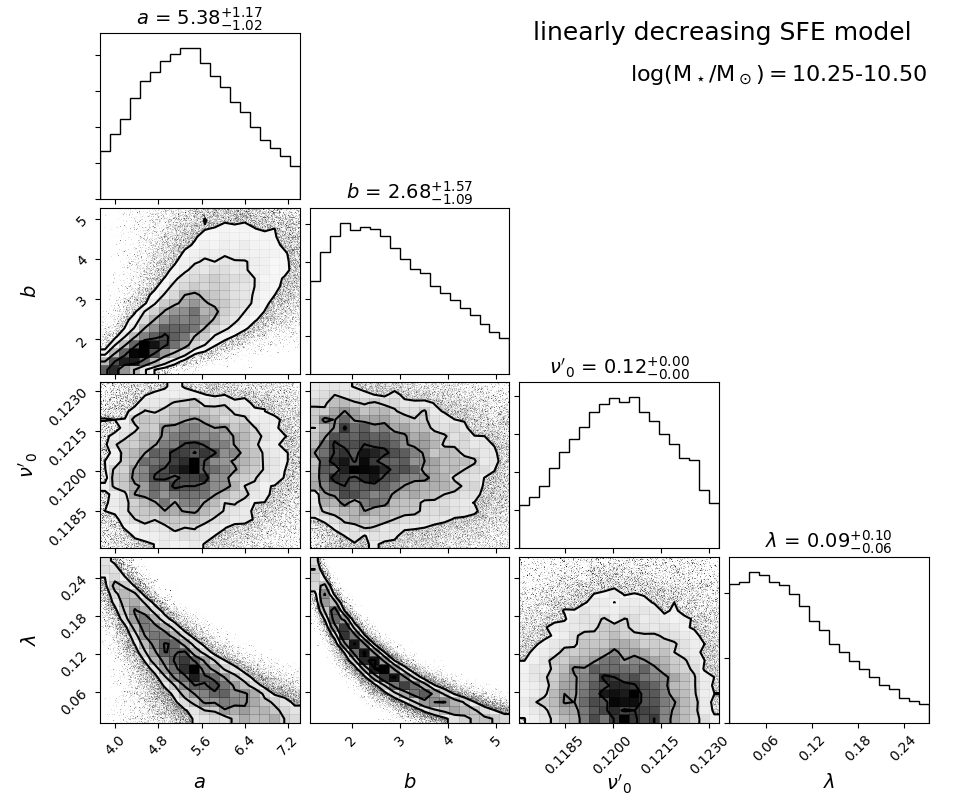}
	\includegraphics[width=0.37\textwidth, trim=0 0 0 0, clip]{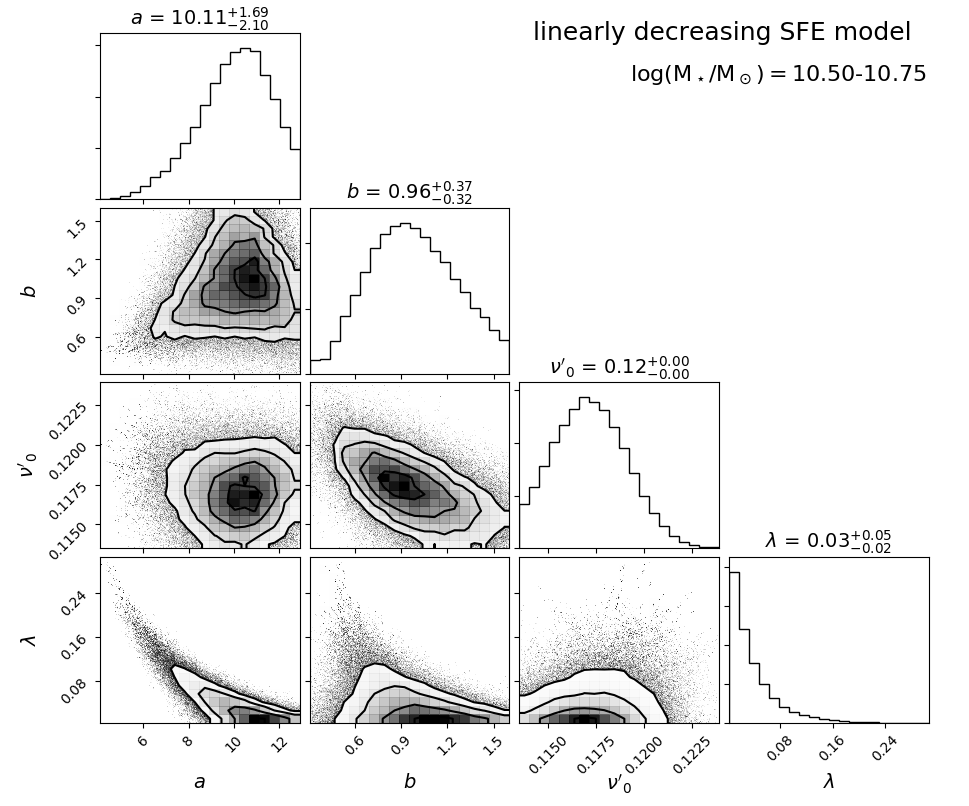}
	\includegraphics[width=0.37\textwidth, trim=0 0 0 0, clip]{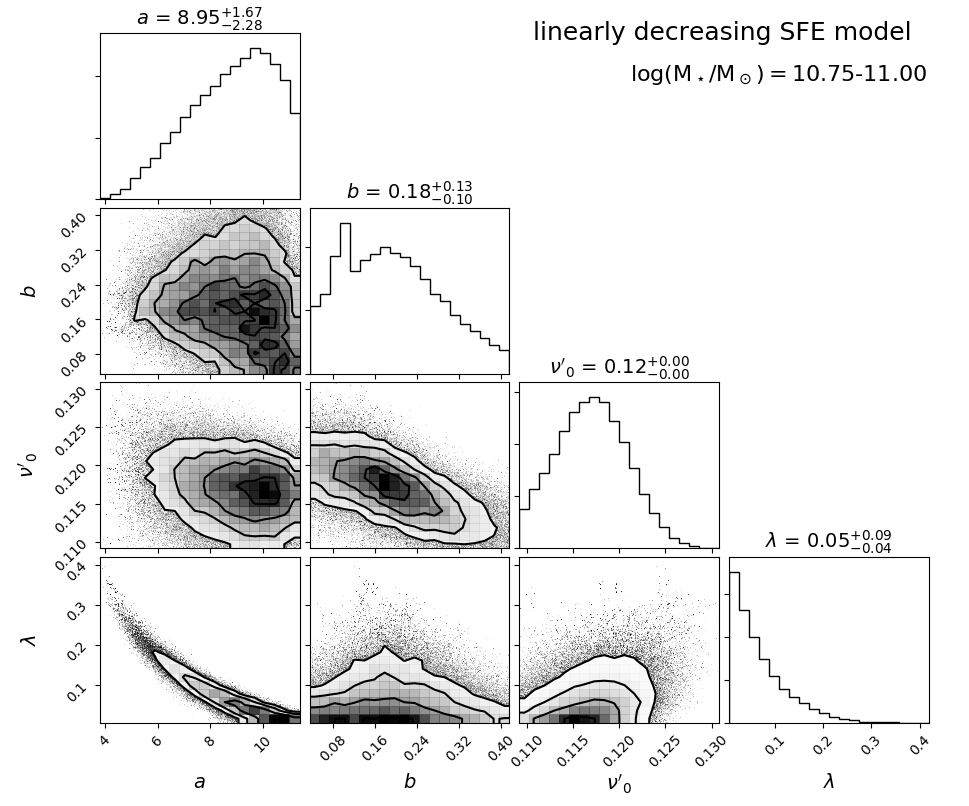}
	
	\caption{Corner plot showing the posterior PDFs of the four model parameters ($a$, $b$, $\nu'_0$, and $\lambda$) for the stacked metallicity gradient for all mass bins, fitted with the linearly decreasing SFE model. The median, 16th and 84th percentiles of the posterior PDF for each parameter are shown above each marginalised PDF.} 
	\label{fig:AppB1}
\end{figure*}

In this appendix we present corner plots showing the posterior PDFs for all mass bins using the $\rm \nu \propto V/r $ (Fig. \ref{fig:AppB}) and linearly decreasing SFE model (Fig. \ref{fig:AppB1}).

%

\section*{Acknowledgements}
\begin{small}
We thank the referee for thought-provoking and constructive comments. F.V. acknowledges support from the European Research Council Consolidator Grant funding scheme (project ASTEROCHRONOMETRY, G.A. n. 772293) and funding from the United Kingdom (UK) Science and Technology Facility Council (STFC) through grant ST/M000958/1. R.M. acknowledge funding from the United Kingdom Science and Technology Facilities Council (STFC) and the European Research Council (ERC), Advanced Grant 695671 `QUENCH'. F.V. thanks the Cavendish Astrophysics Group at the University of Cambridge for kindly supporting his visits in October 2016 and March 2017.
	
	This work makes use of public data from SDSS-IV. Funding for SDSS has been provided by the Alfred P.~Sloan Foundation and Participating Institutions. Additional funding towards SDSS-IV has been provided by the U.S. Department of Energy Office of Science. SDSS-IV acknowledges support and resources from the Centre for High-Performance Computing at the University of Utah. The SDSS web site is {\tt www.sdss.org}.
	
	SDSS-IV is managed by the Astrophysical Research Consortium for the Participating Institutions of the SDSS Collaboration including the  Brazilian Participation Group, the Carnegie Institution for Science, Carnegie Mellon University, the Chilean Participation Group, the French Participation Group, Harvard-Smithsonian Center for Astrophysics, Instituto de Astrof\'isica de Canarias, The Johns Hopkins University, Kavli Institute for the Physics and Mathematics of the Universe (IPMU) / University of Tokyo, Lawrence Berkeley National Laboratory, Leibniz Institut f\"ur Astrophysik Potsdam (AIP),  Max-Planck-Institut f\"ur Astronomie (MPIA Heidelberg), Max-Planck-Institut f\"ur Astrophysik (MPA Garching), Max-Planck-Institut f\"ur Extraterrestrische Physik (MPE), National Astronomical Observatory of China, New Mexico State University, New York University, University of Notre Dame, Observat\'ario Nacional / MCTI, The Ohio State University, Pennsylvania State University, Shanghai Astronomical Observatory, United Kingdom Participation Group, Universidad Nacional Aut\'onoma de M\'exico, University of Arizona, University of Colorado Boulder, University of Oxford, University of Portsmouth, University of Utah, University of Virginia, University of Washington, University of Wisconsin, Vanderbilt University, and Yale University.
	
	The MaNGA data used in this work is publicly available at {\tt http://www.sdss.org/dr13/manga/manga-data/}.

\end{small}


\bibliography{bib22}
\bibliographystyle{mnras}

\bsp

\label{lastpage}

\end{document}